\documentclass{ws-procs961x669}   

\makeatletter
\renewcommand{\ps@plain}{%
\renewcommand{\@oddhead}{\hfil{\footnotesize%
A contribution to the Julian Schwinger Centennial Conference, %
7--12 February 2018, Singapore}\hfil}%
\renewcommand{\@evenhead}{\@oddhead}%
\renewcommand{\@oddfoot}{\hfil\thepage}%
\renewcommand{\@evenfoot}{\thepage\hfil}%
}
\makeatother
\crop[off]

\begin{document}
\title{The magnetoelectric coupling in Electrodynamics}
\author{A. Mart\'\i n-Ruiz}
\address{Instituto de Ciencia de Materiales de Madrid, CSIC \\
Cantoblanco, 28049 Madrid, Spain\\}
{\address{Centro de Ciencias de la Complejidad, Universidad Nacional Aut\'onoma de M\'exico \\
04510 M\'exico, Ciudad de M\'exico, M\'exico \\
E-mail: alberto.martin@nucleares.unam.mx\\
}
\author{M. Cambiaso}
\address{Universidad Andres Bello, Departamento de Ciencias F\'\i  sicas\\
Facultad de Ciencias Exactas, Avenida Rep\'ublica 220, Santiago, Chile\\
E-mail: mcambiaso@unab.cl}
\author{\underline{L. F. Urrutia}}
\address{Instituto de Ciencias Nucleares, Universidad Nacional Aut\'onoma de M\'exico\\
04510 M\'exico, Ciudad de M\'exico, M\'exico\\
E-mail: urrutia@nucleares.unam.mx}

\begin{abstract}
We explore a model akin to axion electrodynamics in which the axion field $\theta (t,\mathbf{x})$ rather than being dynamical   is a piecewise constant effective parameter $\theta$ encoding the microscopic properties of the medium inasmuch as its permittivity or permeability, defining what we call a $\theta$-medium. This model describes a large class of phenomena, among which we highlight the electromagnetic response of materials with topological order, like topological insulators for example. We pursue a Green's function formulation of what amounts to typical boundary-value problems of $\theta$-media, when external sources or boundary conditions are given. As an illustration of our methods, which we have also extended to ponderable media, we  interpret the constant $\theta$ as a novel topological property of vacuum, a so called $\theta$-vacuum, and restrict our discussion to the cases where the permittivity and the  permeability of the media is one. In this way we concentrate upon the effects of the additional $\theta$ coupling which induce remarkable magnetoelectric effects. The issue of boundary conditions for electromagnetic radiation is crucial for the occurrence of the Casimir effect, therefore we apply the methods described above as an alternative way to approach the modifications to the Casimir effect by the inclusion of topological insulators.
\end{abstract}

 
\bodymatter

\section{ Introduction}
\label{INTRO}

Electrodynamics, both the classical \cite{MAXWELL} and the quantum \cite{SCHW1,SCHW2,SCHW3} theories, encompass all our understanding of the interaction between matter and radiation. Although the foundations for the classical theory were laid more than a century ago, still today it is a fruitful research discipline and an excellent arena with potential for new discoveries. Specially when precision measurements are at hand and also when new materials come into play whose novel properties, of ultimate quantum origin, result in new possible forms of interaction between light and such materials. That is the case with topological insulators, as well as other materials with topological order. Interestingly enough, the interaction between matter characterized by topological order, topological insulators among them, and external electromagnetic fields can be described by an extension of Maxwell's theory. In fact, 
in electrodynamics there is the possibility of writing two quadratic gauge
and Lorentz invariant terms: the first one is the usual electromagnetic
density $\mathcal{L}_{\mathrm{EM}}=(\mathbf{E}^{2}-\mathbf{B}^{2})/8\pi $
which yields Maxwell's equations, and the second one is the magnetoelectric
term $\mathcal{L}_{\theta }=\theta \,\mathbf{E}\cdot \mathbf{B}$, where $%
\theta $ is a coupling field usually termed the  axion angle. Many of the
interesting properties of the latter can be recognized from its covariant form $\mathcal{L}_{\theta }=-(\theta /8)\epsilon ^{\mu \nu
\rho \lambda }F_{\mu \nu }F_{\rho \lambda }$, where $\epsilon ^{\mu \nu \rho
\lambda }$ is the Levi-Civita symbol and $F_{\mu \nu }$ is the
electromagnetic field strength. When $\theta $ is globally constant, the $%
\theta $-term is a total derivative and has no effect on Maxwell's
equations. These properties qualify $\mathcal{P}=-(1/8)\epsilon ^{\mu \nu \rho
\lambda }F_{\mu \nu }F_{\rho \lambda } $ to be a topological invariant.
Actually, $\mathcal{P}$ is the simplest example of a Pontryagin density \cite%
{PONT}, corresponding to the abelian group $U(1)$. This structure together
with its generalization to nonabelian groups, has been relevant in diverse
topics in high energy physics such as anomalies \cite{ANOM}, the strong CP
problem \cite{SCP}, topological field theories \cite{TFT} and axions \cite%
{AXIONS}, for example. Recently, an additional application of the
Pontryagin  extended electrodynamics (defined by the full action $\mathcal{L%
}_{\mathrm{EM}}+\mathcal{L}_{\theta }$) has been highlighted in condensed
matter physics, where a piecewise constant axion angle $\theta $ provides an
effective field theory describing the electromagnetic response of a
topological insulator ($\theta =\pi $) in contact with a trivial insulator ($%
\theta =0$) \cite{TI}. A constant $\theta $ can be thought as an additional
parameter characterizing the material in a way analogous to the dielectric
permittivity $\varepsilon $ and the magnetic permeability $\mu $, which
nevertheless manifest only in the presence of a boundary where its value
suddenly changes. 

In this contribution we discuss some general features arising from 
adding to Maxwell's electrodynamics the coupling of the 
Pontryagin density to the scalar field $\theta$,  
leading to a theory that we call $\theta$-electrodynamics ($\theta $-ED),
retaining the name of axion-electrodynamics for the case where the axion
field $\theta$ becomes dynamical. We call the piecewise constant  parameter $\theta$  the magnetoelectric
polarizability (MEP). The resulting field equations  have a wide range of applications in physics. For
example, they describe:  (i)  the electrodynamics of magnetoelectric media \cite{LODELL}, (ii) the
electrodynamics of metamaterials when $\theta $ is a purely complex function %
\cite{MetaMat}, (iii) the electromagnetic response of topological insulators (TIs)
when $\theta =(2n+1)\pi$, with $n$\ integer \cite{TI} and (iv) the
electromagnetic response of Weyl semimetals which can be described by choosing  $\theta (\mathbf{x},t)=2%
\mathbf{b}\cdot \mathbf{x}-2 b_{0}t \, $ \cite{WS}.
Recently, the study of topological insulating and Weyl semimetal phases
either from a theoretical or an experimental perspective has been actively
pursued \cite{TI1, WS1}. 

One of the most remarkable consequences $\theta$-ED is the appearance of the
magnetoelectric effect whereby electric fields induce magnetic fields and vice versa, even for static fields. 
This effect was predicted in
Ref.~\citenum{DZYA} (1959) and subsequently observed in Ref.~\citenum{ASTROV} (1960). For an updated review  of this  effect see for example the Ref.~\citenum{FIEBIG}. A universal topological magnetoelectric effect has recently been measured in TIs \cite{METI}.
Many additional  interesting
magnetoelectric effects arising from $\theta $-ED have been highlighted
using different approaches. For example, electric charges close to the
interface between two $\theta$-media induce image  magnetic monopoles (and vice versa) \cite%
{science,Kim1,Kim2,Wilczek}. Also, the propagation of electromagnetic waves across
a $\theta $-boundary have been studied finding that a non trivial Faraday
rotation of the polarizations appears \cite{Kim1, Kim2, ZH, Hehl}. The shifting of the spectral lines in  hydrogen-like 
ions placed in front of a planar TI, as well as the modifications to the Casimir Polder potential in the non-retarded approximation were studied in Ref.~\citenum{PRA}. The classical dynamics of a Rydberg hydrogen atom near a planar TI has also been investigated \cite{EPL}.

The paper is organized as follows. In section \ref{thetamedium} we present a brief review of electrodynamics in 
media characterized by a parameter $\theta$ (to be called a $\theta$-medium), recalling their most important properties.  Section \ref{Green} contains a
summary of our generalized Green's function method to construct the corresponding electromagnetic fields
produced by  charges, currents and boundary conditions in systems subjected to the following coordinate conditions: (i) the coordinates can be chosen  in such a way that
the interface between two media with different values  $\theta$ is  defined by setting constant only one of them and (ii) the Laplacian is separable in such coordinates.  The particularly simple case of planar symmetry is discussed subsequently in section \ref{PLS}, where the reader is also referred to the analogous  extensions to cylindrical and spherical coordinates.
As a specific application of our methods to the case of a planar interface, the Casimir effect between two metallic plates with a topological insulator between them is considered in section \ref{CASE}. Our conventions are taken from Ref.~\citenum{Jackson}, where   $F_{\mu\nu}=\partial_\nu A_\nu-\partial_\nu A_\mu$, ${\tilde F}^{\mu\nu}=\epsilon^{\mu\nu\alpha\beta} F_{\alpha\beta}/2$  $%
F^{i0}=E^{i}$, $F^{ij}=-\epsilon^{ijk}B^{k}$ and $\tilde{F}^{i0}=B^{i}$, 
$\tilde{F}^{ij}=\epsilon ^{ijk}E^{k}$. Also $\mathbf{V}=(V^i)=(V_x, V_y, V_z )$ for any vector $\mathbf{V}$. The metric is $(+,-,-,-)$ and $\epsilon^{0123}=+1=\epsilon^{123}$

\section{ Electrodynamics in a $\protect\theta $-medium}
\label{thetamedium}
Electromagnetic phenomena in material media are described by the Maxwell's
field equations, 
\begin{equation}
\nabla \cdot \mathbf{D}=4\pi \rho ,\quad \nabla \cdot \mathbf{B}=0,\quad
\nabla \times \mathbf{E}+\frac{1}{c}\frac{\partial \mathbf{B}}{\partial t}%
=0,\quad \nabla \times \mathbf{H}-\frac{1}{c}\frac{\partial \mathbf{D}}{%
\partial t}=\frac{4\pi }{c}\mathbf{J},  \label{MAXMM}
\end{equation}%
together with constitutive relations giving the displacement $\mathbf{D}$
and the magnetic field $\mathbf{H}$ in terms of the electric $\mathbf{E}$
and magnetic induction $\mathbf{B}$ fields, plus the Lorentz force \cite{Jackson}. These depend on
the nature of the material, and they are generally of the form $\mathbf{D}=%
\mathbf{D}(\mathbf{E},\mathbf{B})$ and $\mathbf{H}=\mathbf{H}(\mathbf{E},%
\mathbf{B})$. For instance, for linear media they are $\mathbf{D}%
=\varepsilon \mathbf{E}$ and $\mathbf{H}=\mathbf{B}/\mu $, where $%
\varepsilon $ is the dielectric permittivity and $\mu $ is the magnetic
permeability. For isotropic materials $\varepsilon $ and $\mu $ are
constants, while for anisotropic materials they are tensorial in nature and
may depend on the spacetime coordinates.

In this paper we are concerned with a particular class of materials
described by the following constitutive relations 
\begin{equation}
\mathbf{D}=\varepsilon \mathbf{E}-\frac{\theta \alpha }{\pi }\mathbf{B}%
,\qquad \mathbf{H}=\frac{1}{\mu }\mathbf{B}+\frac{\theta \alpha }{\pi }%
\mathbf{E},  \label{ec3}
\end{equation}%
where $\alpha \simeq 1/137$ is the fine structure constant and the MEP $\theta $
is an additional parameter of the medium, which can be considered on
the same footing as the permittivity $\varepsilon $ or the permeability 
$\mu$. In the general situation these parameters may be functions of the
spacetime coordinates. The constitutive relations \ (\ref{ec3}) yield the
following inhomogeneous Maxwell's equations 
\begin{equation}
\nabla \cdot (\varepsilon \mathbf{E})=4\pi \rho +\frac{\alpha }{\pi }\nabla
\theta \cdot \mathbf{B},\quad \nabla \times \left( \mathbf{B}/\mu \right) -%
\frac{1}{c}\frac{\partial (\varepsilon \mathbf{E})}{\partial t}=\frac{4\pi }{%
c}\mathbf{J}-\frac{\alpha }{\pi }\nabla \theta \times \mathbf{E}-\frac{1}{c}%
\frac{\alpha }{\pi }\frac{\partial \theta }{\partial t}\mathbf{B}.
\label{MAXMOD}
\end{equation}%
In fact, the modified Maxwell's equations (\ref{MAXMOD}) can be derived from
the usual electromagnetic action supplemented with the coupling of the
abelian Pontryagin density $\mathcal{P}$
via the MEP $\theta$   
\begin{equation}
S[ \Phi, \textbf{A} ] = \int dt \, d ^{3} \textbf{x} \left[ \frac{1}{8 \pi} \left( \varepsilon \textbf{E} ^2 - \frac{1}{\mu} \textbf{B} ^2 \right ) - \frac{\alpha}{4 \pi ^{2}} \theta(\textbf{x}) \, \textbf{E} \cdot \textbf{B} - \rho \Phi 
+ \frac{1}{c} \textbf{J} \cdot \textbf{A} \right]. \label{action}
\end{equation}
The electromagnetic fields $\mathbf{E}$ and $%
\mathbf{B}$ are written in term of  the electromagnetic potentials $\Phi $ and $%
\mathbf{A}$ as usual, providing  a solution of the homogeneous equations in Eq. (\ref{MAXMM}), which are summarized in the Bianchi identity $\partial_\mu {\tilde F}^{\mu\nu}=0$ 

An important consequence of the modified Maxwell's equations (\ref{MAXMOD})
is the appearance of additional field-dependent effective charge and current
densities given by 
\begin{equation}
\rho _{\theta }=\frac{\alpha }{4\pi ^{2}}\nabla \theta \cdot \mathbf{B}%
,\qquad \mathbf{J}_{\theta }=-\frac{c\alpha }{4\pi ^{2}}\nabla \theta \times 
\mathbf{E}-\frac{\alpha }{4\pi ^{2}}\frac{\partial \theta }{\partial t}%
\mathbf{B}.  \label{DENEFF}
\end{equation}%
Current conservation $\nabla \cdot \mathbf{J}_{\theta }+\partial \rho
_{\theta }/\partial t=0$ can be directly verified as a consequence of the
homogeneous equations in (\ref{MAXMM}). Note that these expressions depend
only on spacetime gradients of the MEP $\theta $. This is because the
Pontryagin density $\mathcal{P}$ is a total derivative in such way that the
coupling in\ (\ref{action}) does not affect the equations of motion when $%
\theta$ is globally a constant. Even though the constitutive relations
depend upon the constant $\theta $, their contribution to the equations of
motion turns out to be null due to the homogeneous Maxwell's equations.
This can be directly verified from the constitutive relations (\ref{ec3}),
yielding 
\begin{eqnarray}
\nabla \cdot \mathbf{D} &=&\nabla \cdot \left( \varepsilon \mathbf{E}\right)
-\frac{\theta \alpha }{\pi }\left( \nabla \cdot \mathbf{B}\right) \nonumber ,\qquad \\
\nabla \times \mathbf{H-}\frac{1}{c}\frac{\partial \mathbf{D}}{\partial t}
&=&\nabla \times \left( \frac{1}{\mu }\mathbf{B}\right) -\frac{1}{c}\frac{%
\partial }{\partial t}\left( \varepsilon \mathbf{E}\right) +\frac{\theta
\alpha }{\pi }\left( \nabla \times \mathbf{E}+\frac{1}{c}\frac{\partial }{%
\partial t}\mathbf{B}\right). \label{VERCR}
\end{eqnarray}

Physically, the effective charge and current densities (\ref{DENEFF}) encode one of the most 
remarkable properties of $\theta$-ED, which is the magnetoelectric effect.

A large class of interesting phenomena can be described by $\theta $-ED if one considers the adjacency of different media with constant $\theta $.  In
the simplest case where the $(3+1)$-dimensional spacetime is $\mathcal{M}=%
\mathcal{U}\times \mathbb{R}$, with $\mathcal{U}$ being a three-dimensional
manifold and $\mathbb{R}$ corresponding to the temporal axis, we make a
partition of space in two regions: $\mathcal{U}_{1}$ and $\mathcal{U}_{2}$,
in such a way that manifolds $\mathcal{U}_{1}$ and $\mathcal{U}_{2}$
intersect along a common two-dimensional boundary $\Sigma $, to be called
the $\theta $-boundary, so that $\mathcal{U}=\mathcal{U}_{1}\cup \mathcal{U}%
_{2}$ and $\Sigma =\mathcal{U}_{1}\cap \mathcal{U}_{2}$, as shown in Fig. %
\ref{regions}.

 We also assume that the MEP $\theta $ is piecewise constant in
such way that it takes the value $\theta =\theta _{1}$ in the region $%
\mathcal{U}_{1}$ and the  value $\theta =$ $\theta _{2}$ in the region $%
\mathcal{U}_{2}$. This situation is expressed in the characteristic function 
\begin{equation}
\theta \left( \mathbf{x}\right) =\left\{ 
\begin{array}{c}
\theta _{1}\;\;\;,\;\;\;\mathbf{x}\in \mathcal{U}_{1} \\ 
\theta _{2}\;\;\;,\;\;\;\mathbf{x}\in \mathcal{U}_{2}%
\end{array}%
\right. .  \label{theta}
\end{equation}%
The two-dimensional surface $\Sigma $\ is parametrized by
some function $F_{\Sigma }(\mathbf{x})=0$, such that 
\begin{equation}
n_{\mu }=(0,\mathbf{\hat{n})=\partial }_{\mu }F_{\Sigma }(\mathbf{x}),
\label{UNITNORMAL}
\end{equation}%
is the outward unit normal to $%
\Sigma $ with respect to the region $\mathcal{U}_{1}$.
In this scenario the $\theta $-term in the action fails to be a global  total derivative
 because it is defined over a region with the boundary $%
\Sigma $. Consequently the modified Maxwell's
equations acquire additional effective charge and current densities with support only at
the boundary (in the following we set $c=1$)  
\begin{figure}[tbp]
\begin{center}
\includegraphics [width=3in]
{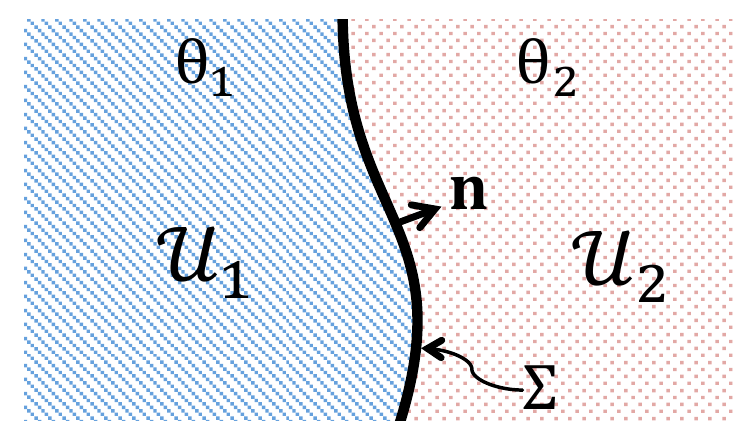}
\end{center}
\caption{{\protect\small Region over which the electromagnetic field theory
is defined.}}
\label{regions}
\end{figure}
\begin{eqnarray}
\nabla \cdot \mathbf{E} &=&\tilde{\theta}\delta \left( F_{\Sigma }(\mathbf{x}%
)\right) \mathbf{B}\cdot \mathbf{\hat{n}}+4\pi \rho ,  \label{GaussE} \\
\nabla \times \mathbf{B}-\frac{\partial \mathbf{E}}{\partial t} &=&\tilde{%
\theta}\delta \left( F_{\Sigma }(\mathbf{x})\right) \mathbf{E}\times \mathbf{%
\hat{n}}+4\pi \mathbf{J},  \label{Ampere}
\end{eqnarray}%
which reproduce the Eqs. (\ref{MAXMOD}) in this setting. The homogeneous
equations are included in the Bianchi identity. Here $\mathbf{\hat{n}}$ is
the unit normal to $\Sigma $  defined in Eq. (\ref{UNITNORMAL}),  shown
in  Fig.~\ref{regions} and $\tilde{\theta}=\alpha \left( \theta _{1}-\theta _{2}\right) /\pi $,
which enforces the invariance of the classical action under the shifts of $%
\theta $ by any constant, $\theta \rightarrow \theta +C$. As we see from Eqs. (\ref{GaussE}-\ref{Ampere})
the behavior of $\theta $-ED in the bulk regions $\mathcal{U}_{1}$ and $%
\mathcal{U}_{2}$ is the same as in standard electrodynamics. 

Assuming that the time derivatives of the fields are finite in the vicinity
of the surface $\Sigma$, the field equations (\ref{GaussE}) and (\ref{Ampere}) imply that the normal component
of $\mathbf{E}$, and the tangential components of $\mathbf{B}$, acquire
discontinuities additional to those produced by superficial free charges and
currents, while the normal component of $\mathbf{B}$, and the tangential
components of $\mathbf{E}$, are continuous at the boundary. For vanishing external sources
on $\Sigma $ the boundary conditions read: 
\begin{eqnarray}
\Delta \mathbf{E}_{n}\big|_{\Sigma } &=&\tilde{\theta}\mathbf{B}_{n}\big|%
_{\Sigma }, \quad  \Delta \mathbf{B}_{\parallel }\big|_{\Sigma } =-\tilde{\theta}\mathbf{E}%
_{\parallel }\big|_{\Sigma },   \label{BC1} \\
\Delta \mathbf{B}_{n}\big|_{\Sigma } &=&0,  \quad 
\Delta \mathbf{E}_{\parallel }\big|_{\Sigma }=0.  \label{BC2}
\end{eqnarray}
The notation $\Delta \mathbf{V}_{i}\big|_{\Sigma}$ refers to the
discontinuity of the $i$-th component of the vector $\mathbf{V}$ across the
interface $\Sigma$, while $\mathbf{V}_{j}\big|_{\Sigma}$ indicates the
continuous value of the $j$-th component evaluated at $\Sigma$. The continuity
conditions, (\ref{BC2}), imply that the right
hand sides of equations (\ref{BC1})  are well
defined and they represent surface charge and current densities,
respectively. An immediate consequence of the boundary conditions (\ref%
{BC1}) and (\ref{BC2}) is that the presence of a magnetic field crossing the surface $\Sigma$
is sufficient to generate an electric surface charge density there,
even in the absence of free electric charges.

\section{The Green's function method in a $\theta$-vacuum}

\label{Green}

In this section we review the  Green's function (GF) method to solve a class of static
boundary-value problems in $\theta $-ED in terms of the electromagnetic
potential $A^{\mu }$. Certainly one could solve for the electric and
magnetic fields from the modified Maxwell equations together with the
boundary conditions (\ref{BC1}-\ref{BC2}), however, just as in
ordinary electrodynamics, there might be occasions where information about
the sources is unknown and rather we are provided with information of the
4-potential at some given boundaries. In these cases, the GF method provides
the general solution to such boundary-value problem (Dirichlet or
Neumann) for arbitrary sources. Nevertheless, in the following we restrict
ourselves to contributions of free sources only outside the $\theta$-boundary with 
no additional  boundary conditions (BCs) besides  those required at $\Sigma$. Also we consider
the simplest media having $\theta_1\neq \theta_2$, but  with $\varepsilon=1$ and $\mu=1$, which we call the $\theta$-vacuum.

In this case, the inhomogeneous Maxwell's equations can be written as
\begin{equation}
\partial _{\mu }F^{\mu \nu }=\tilde{\theta}\delta \left( F_{\Sigma }(\mathbf{%
x})\right) n_{\mu }\tilde{F}^{\mu \nu }+4\pi j^{\nu },  \label{FieldEqs}
\end{equation}%
Current
conservation can be verified directly by taking the divergence on both sides
of Eq.~(\ref{FieldEqs}) and realizing that 
\begin{equation}
\partial _{\nu }\left( \tilde{\theta}\delta \left( F_{\Sigma }(\mathbf{x}%
)\right) n_{\mu }\tilde{F}^{\mu \nu }\right) =\tilde{\theta}\delta ^{\prime
}\left( F_{\Sigma }(\mathbf{x})\right) n_{\nu }n_{\mu }\tilde{F}^{\mu \nu }+%
\tilde{\theta}\delta \left( F_{\Sigma }(\mathbf{x})\right) n_{\mu }\partial
_{\nu }\tilde{F}^{\mu \nu }
\end{equation}
is zero by symmetry properties together with the Bianchi identity.
Since the homogeneous Maxwell equations are not modified, the electrostatic and
magnetostatic fields can be written in terms of the $4$-potential $A^{\mu
}=\left( \phi ,\mathbf{A}\right) $ according to $\mathbf{E}=-\nabla \phi $
and $\mathbf{B}=\nabla \times \mathbf{A}$ as usual. In the Coulomb gauge $%
\nabla \cdot \mathbf{A}=0$, the $4$-potential satisfies the equation of
motion 
\begin{equation}
\left[ -\eta _{\phantom{\mu}\nu }^{\mu }\nabla ^{2}-\tilde{\theta}\delta \left(
F_{\Sigma }(\mathbf{x})\right) n_{\rho }\epsilon _{\;\;\;\;\ \nu }^{\rho \mu
\alpha }\partial _{\alpha }\right] A^{\nu }=4\pi j^{\mu },
\label{FiedlEqPlaneConfig}
\end{equation}%
together with the boundary conditions 
\begin{equation}
\Delta A^{\mu }\big|_{\Sigma }=0, \qquad  \Delta \left( \partial
_{z}A^{\mu }\right) \big|_{\Sigma }=-\tilde{\theta}\epsilon _{\;\;\;\;\ \nu
}^{3\mu \alpha }\left( \partial _{\alpha }A^{\nu }\right) \big|_{\Sigma }.
\label{BC4Pot}
\end{equation}%
One can further check that these boundary conditions for the $4$-potential
correspond to those written  in Eqs. (\ref{BC1}-\ref{BC2}).

To obtain a general solution for the potentials $\phi $ and $\mathbf{A}$ in
the presence of arbitrary external sources $j^{\mu }(\mathbf{x})$, we
introduce the GF $G_{\phantom{\nu}\sigma }^{\nu }\left( \mathbf{x},\mathbf{x}^{\prime
}\right) $ solving Eq. (\ref{FiedlEqPlaneConfig}) for a point-like source, 
\begin{equation}
\left[ -\eta _{\phantom{\mu}\nu }^{\mu }\nabla ^{2}-\tilde{\theta}\delta \left(
F_{\Sigma }(\mathbf{x})\right) n_{\rho }\epsilon _{\phantom{\rho}\phantom{\mu}\phantom{\alpha} \nu }^{\rho \mu
\alpha }\partial _{\alpha }\right] G_{\phantom{\nu}\sigma }^{\nu }\left( \mathbf{x},%
\mathbf{x}^{\prime }\right) =4\pi \eta _{\;\sigma }^{\mu }\delta ^{3}\left( 
\mathbf{x}-\mathbf{x}^{\prime }\right) ,  \label{EqsGreenMatrix}
\end{equation}%
together with the boundary conditions (\ref{BC4Pot}), in such a way that the
general solution for the $4$-potential in the Coulomb gauge is 
\begin{equation}
A^{\mu }\left( \mathbf{x}\right) =\int d^{3}\mathbf{x}^{\prime }\;G_{\phantom{\mu}\nu
}^{\mu }\left( \mathbf{x},\mathbf{x}^{\prime }\right) j^{\nu }\left( \mathbf{%
x}^{\prime }\right) .  \label{GreenMatrix}
\end{equation}%
According to Eqs. (\ref{EqsGreenMatrix}) the diagonal entries of the GF
matrix are related with the electric and magnetic fields arising from the
charge and current density sources, respectively, although they acquire a $%
\theta $-dependence. However, the non-diagonal terms encode the
magnetoelectric effect, \textit{i.e.} the charge (current) density
contributing to the magnetic (electric) field.

As we will show in the following, a  further simplification in $\theta $-ED arises when the system satisfies the following two coordinate conditions:(i) the coordinate system can
be chosen so that the interface $\Sigma$ is defined by setting constant only one
of them and  (ii)  the Laplacian is separable in such coordinates in such a way that a complete orthonormal set of eigenfunctions can be defined in the subspace orthogonal to the coordinate defining the interface. 
Three cases show up
immediately: (i) a plane interface at fixed $z$,  (ii) a spherical interface at
constant $r$  and (iii) a cylindrical interface at
constant $\rho $. In all this
cases the characteristic function $\theta \left( \mathbf{x}\right) \;$%
defined in Eq. (\ref{theta}) \ can be written in terms of the Heaviside function 
$H$ of one coordinate in terms of  $H(z-a),\; H(r-a)$ and $H(\rho -a)$,
with the associated unit vectors $\mathbf{\hat{n}}_{\xi }\;$given by \ $%
\mathbf{\hat{k}}$\ ,\ $\mathbf{\hat{r}\;}$and $\mathbf{\hat{\rho}}$,
respectively, in each of the adapted coordinate systems. Then Eq. (\ref%
{EqsGreenMatrix}) reduces to
\begin{equation}
\left[ -\eta _{\phantom{\mu}\nu }^{\mu }\nabla ^{2}-\tilde{\theta}\delta \left( \xi
-\xi _{0}\right) \epsilon _{\phantom{\xi}\phantom{\mu}\phantom{\alpha} \nu }^{\xi \mu \alpha }\partial
_{\alpha }\right] G_{\phantom{\nu}\sigma }^{\nu }\left( \mathbf{x},\mathbf{x}^{\prime
}\right) =4\pi \eta _{\phantom{\mu}\sigma }^{\mu }\delta ^{3}\left( \mathbf{x}-\mathbf{x%
}^{\prime }\right) ,  \label{GFEQ}
\end{equation}%
where $\xi$ denotes the  coordinate defining the interface at 
$\xi =\xi _{0}$ and the coupling of the $\theta $-term\ is given by a one
dimensional delta function with support only in the coordinate that defines the interface. Also, the unit vector $%
\mathbf{\hat{n}}_{\xi }$ will have a component only in the direction $\xi$.

Let us consider the coordinates partitioned according to $\xi $ plus two
additional ones which we denote by $\sigma $ and $\tau $. Also  assume 
that the Laplacian can be separated in the form
\begin{equation}
\nabla ^{2}=L_{1}(\xi )+f(\xi )L_{2}(\sigma ,\tau )  \label{NABLA2}
\end{equation}%
where the operator $L_{2}(\sigma ,\tau )\;$has eigenfunctions $\Psi
_{M}(\sigma ,\tau )\;$which form a complete orthonormal set in the subspace
of the coordinates $\sigma ,\tau $ (which we denote collectively by $\Pi $)%
\begin{equation}
L_{2}(\sigma ,\tau )\Psi _{M}(\sigma ,\tau )=\lambda _{M}\Psi _{M}(\sigma
,\tau ), \label{EIGENF}
\end{equation}%
where $M$  denote a set of discrete or continuous labels. The basic
properties of \ $\Psi _{M}(\sigma ,\tau )$ are  
\begin{eqnarray}
\int d\mu (\sigma ,\tau )\;\Psi _{M}^{\ast }(\sigma ,\tau )\Psi _{M^{\prime
}}(\sigma ,\tau ) =\delta _{M,M^{\prime }},\;\; && \;\;\sum_{M}\Psi
_{M}(\sigma ,\tau )\Psi _{M}^{\ast }(\sigma ^{\prime },\tau ^{\prime
})=\delta ^{2}(\Pi -\Pi ^{\prime }),\;  \notag \\
\int d\mu (\sigma ,\tau )\delta ^{2}(\Pi -\Pi ^{\prime })=1 &&, \label{ORTCOMP}
\end{eqnarray}%
where $d\mu$ denotes the integration measure  in each subspace and $d^{3}%
\mathbf{x}=d\mu (\sigma,\tau )\;d\mu (\xi )$. Also we have   
\begin{equation*}
\;\delta ^{3}\left( \mathbf{x}-\mathbf{x}^{\prime }\right) =\delta ^{2}(\Pi
-\Pi ^{\prime })\delta (\xi -\xi ^{\prime }),\;\;\;\int d\mu (\xi )\delta
(\xi -\xi ^{\prime })=1. \label{DELTAREP}
\end{equation*}%
Next we introduce the reduced Green's function$\;\;\left( g_{\;\sigma }^{\nu
}\right) _{M,M^{\prime }}\left( \xi ,\xi ^{\prime }\right) \;$in the
following way 
\begin{equation}
G_{\phantom{\nu}\sigma }^{\nu }\left( \mathbf{x},\mathbf{x}^{\prime }\right)
=4\pi \sum_{M,M^{\prime }}\Psi _{M}(\sigma ,\tau )\Psi _{M^{\prime }}^{\ast
}(\sigma ^{\prime },\tau ^{\prime })\left( g_{\phantom{\nu}\sigma }^{\nu }\right)
_{M,M^{\prime }}\left( \xi ,\xi ^{\prime }\right)   \label{FULLGF}
\end{equation}%
When substituting Eq. (\ref{NABLA2}) in Eq. (\ref{GFEQ}) we obtain 
\begin{eqnarray}
&&\sum_{M,M^{\prime }}\Psi _{M}(\sigma ,\tau )\Psi _{M^{\prime }}^{\ast
}(\sigma^{\prime },\tau ^{\prime })\left[ -\eta _{\phantom{\mu}\nu }^{\mu }L_{1}(\xi )%
\right] \left( g_{\phantom{\nu}\sigma }^{\nu }\right) _{M,M^{\prime }}\left( \xi ,\xi
^{\prime }\right) \nonumber   \\
&&+\sum_{M,M^{\prime }}\Psi _{M}(\sigma ,\tau )\Psi _{M^{\prime }}^{\ast
}(\sigma ^{\prime },\tau ^{\prime })\left[ -\eta _{\phantom{\mu}\nu }^{\mu }f(\xi
)\lambda _{M}\right] \left( g_{\phantom{\nu}\sigma }^{\nu }\right) _{M,M^{\prime
}}\left( \xi ,\xi ^{\prime }\right) \nonumber  \\
&&+\sum_{M,M^{\prime }}O_{\phantom{\mu}\nu }^{\mu }(\sigma ,\tau )\Psi _{M}(\sigma
,\tau )\Psi _{M^{\prime }}^{\ast }(\sigma ^{\prime },\tau ^{\prime })\left[ -%
\tilde{\theta}\delta \left( \xi -\xi _{0}\right) \right] \left( g_{\phantom{\nu}\sigma
}^{\nu }\right) _{M,M^{\prime }}\left( \xi ,\xi ^{\prime }\right) \nonumber   \\
&=& \eta _{\phantom{\mu}\sigma }^{\mu }\sum_{N,N^{\prime }}\Psi _{N}(\sigma ,\tau
)\Psi _{N^{\prime }}^{\ast }(\sigma ^{\prime },\tau ^{\prime }))\delta
_{N,N^{\prime }}\delta (\xi -\xi ^{\prime }), \label{SEP}
\end{eqnarray}%
since  the operator 
\begin{equation}
\epsilon _{\phantom{\xi}\phantom{\mu}\phantom{\alpha} \nu }^{\xi \mu \alpha }\partial _{\alpha }\equiv
O_{\phantom{\mu}\nu }^{\mu }(\sigma ,\tau ) \label{OP}
\end{equation}%
contains only derivatives with respect to $\sigma ,\tau $ so that  it
acts  upon the functions $\Psi _{M}(\sigma ,\tau )$. Multiplication to the
right by $\Psi _{T}(\sigma ^{\prime },\tau ^{\prime })$ and
integration over $d\mu (\sigma ^{\prime },\tau ^{\prime })$, followed by 
multiplication   to the left by $\Psi _{P}^{\ast }(\sigma ,\tau )$ and
integration over $d\mu (\sigma ,\tau )$ yields 
\begin{eqnarray}
\left[ \left( L_{1}(\xi )+\lambda _{P}f(\xi )\right) \right] \left(
g_{\phantom{\mu}\sigma }^{\mu }\right) _{P,T}\left( \xi ,\xi ^{\prime }\right) && +\tilde{%
\theta}\delta \left( \xi -\xi _{0}\right) \sum_{M}\left[ O_{\phantom{\mu}\nu }^{\mu }%
\right] _{PM}\left( g_{\phantom{\nu}\sigma }^{\nu }\right) _{M,T}\left( \xi ,\xi
^{\prime }\right)
\nonumber \\ 
&&=- \eta _{\phantom{\mu}\sigma }^{\mu }\delta _{P,T}\delta (\xi
-\xi^{\prime }), \label{REDGFFINAL}
\end{eqnarray}%
where we have  introduced the following  matrix element 
\begin{equation}
\left[ O_{\phantom{\mu}\nu }^{\mu }\right] _{PM}\equiv \int d\mu (\sigma ,\tau )\Psi
_{P}^{\ast }(\sigma ,\tau )O_{\phantom{\mu}\nu }^{\mu }(\sigma ,\tau )\Psi _{M}(\sigma
,\tau ) \label{ME1}
\end{equation}%
which is  independent of $\xi$ and $\xi ^{\prime }$.

In this way we transform  the calculation of the reduced GF  into a one
dimensional problem with a delta interaction. The above equation (\ref%
{REDGFFINAL}) can be directly integrated with the knowledge of  an additional reduced
GF $\left( \mathfrak{g}_{\phantom{\mu}\sigma }^{\mu }\right) _{P,T}\left( \xi ,\xi
^{\prime }\right)$, corresponding to the $\tilde{\theta}=0$ limit,
which satisfies  
\begin{equation}
\left[ \left( L_{1}(\xi )+\lambda _{P}f(\xi )\right) \right] \left( 
\mathfrak{g}_{\phantom{\mu}\sigma }^{\mu }\right) _{P,T}\left( \xi ,\xi ^{\prime
}\right) =- \eta _{\phantom{\mu}\sigma }^{\mu }\delta _{P,T}\delta (\xi -\xi
^{\prime }),  \label{FREEREDUCEDGF}
\end{equation}%
plus boundary conditions.  

The introduction of $\left( \mathfrak{g}_{\phantom{\mu}\sigma }^{\mu }\right)
_{P,T}\left( \xi ,\xi ^{\prime }\right) $ derives from the existence of a
full Green's function%
\begin{equation}
\mathfrak{G}_{\phantom{\nu}\sigma }^{\nu }\left( \mathbf{x},\mathbf{x}^{\prime }\right)
=\sum_{M,M^{\prime }}\Psi _{M}(\sigma ,\tau )\Psi _{M^{\prime }}^{\ast
}(\sigma ^{\prime },\tau ^{\prime })\left( \mathfrak{g}_{\phantom{\nu}\sigma }^{\nu
}\right) _{M,M^{\prime }}\left( \xi ,\xi ^{\prime }\right), \label{FULLFREEGF}
\end{equation}%
which  must respect the coordinate conditions of the problem in a setting where 
 the $\theta$-medium is absent. We refer to them as the free GF's,
emphasizing that they correspond to the $\tilde{\theta}=0$ case.   These
GF's can be taken directly from the vast literature in standard
electrodynamics and are the basis for finding the response of an identical
system now in the presence of a $\theta$-medium, the interface of which defines  the corresponding 
coordinate conditions. As an illustration take the case of a planar $\theta $-medium that
 can be embedded in two different ways: (i) either in vacuum, just by choosing the free GF $\mathfrak{G}%
_{\phantom{\nu}\sigma }^{\nu }\left( \mathbf{x},\mathbf{x}^{\prime }\right) $ with
standard BCs at infinity, or (ii)  between  a pair of conducting plates of infinite extension which are  parallel to the interface 
just by requiring $\mathfrak{G}_{\phantom{\nu}\sigma }^{\nu }\left( \mathbf{x},\mathbf{x}%
^{\prime }\right) $ to satisfy the appropriate BCs at the plates, which can be found
in  Ref.~\citenum{CED}, for example. This approach was used in Ref.~\citenum{EPL2}
when calculating the Casimir effect between parallel metallic plates  in the presence of a planar  $\theta $%
-medium and will be reviewed in section \ref{CASE}. 

In terms of the free reduced GF $\left( \mathfrak{g}_{\phantom{\mu}\sigma }^{\mu }\right) _{P,T}\left(
\xi ,\xi ^{\prime }\right) $ we obtain 
\begin{equation}
\left( g_{\phantom{\mu}\sigma }^{\mu }\right) _{P,T}\left( \xi ,\xi ^{\prime }\right)
=\left( \mathfrak{g}_{\phantom{\mu}\sigma }^{\mu }\right) _{P,T}\left( \xi ,\xi
^{\prime }\right) +\frac{\tilde{\theta}}{4\pi }\sum_{M,N}\left( \mathfrak{g}%
_{\phantom{\mu}\rho }^{\mu }\right) _{P,N}\left( \xi ,\xi _{0}\right) \left[ O_{\phantom{\rho}\nu
}^{\rho }\right] _{NM}\left( g_{\phantom{\nu}\sigma }^{\nu }\right) _{M,T}\left( \xi
_{0},\xi ^{\prime }\right).   \label{SOLGFFINAL}
\end{equation}%
This result can be explicitly verified by applying the operator $\left[
\left( L_{1}(\xi )+\lambda _{P}f(\xi )\right) \right] $ to Eq. (\ref%
{SOLGFFINAL}) and using  Eq. (\ref{FREEREDUCEDGF}).

It is convenient to think of $\left( g_{\phantom{\mu}\sigma }^{\mu }\right)
_{P,T},\;\left( \mathfrak{g}_{\phantom{\mu}\sigma }^{\mu }\right) _{P,T}$ and $\left[
O_{\phantom{\rho}\nu }^{\rho }\right] _{NM}$ as generalized matrix elements of the
operators $g,\;\mathfrak{g}$ and $O$, respectively. This allows us to rewrite Eq. (\ref%
{SOLGFFINAL}) in the compact form%
\begin{equation}
g\left( \xi ,\xi ^{\prime }\right) =\mathfrak{g}\left( \xi ,\xi ^{\prime
}\right) +\frac{\tilde{\theta}}{4\pi }\mathfrak{g}\left( \xi ,\xi
_{0}\right) Og\left( \xi _{0},\xi ^{\prime }\right).   \label{COMGFEQ}
\end{equation}
This set of equations constitute a coupled system of algebraic
equations which can be disentangled  according to the following steps. First we set $\xi =\xi _{0}$ in Eq. (\ref{COMGFEQ})
\begin{equation}
g\left( \xi _{0},\xi ^{\prime }\right) =\mathfrak{g}\left( \xi _{0},\xi
^{\prime }\right) +\frac{\tilde{\theta}}{4\pi }\mathfrak{g}\left( \xi
_{0},\xi _{0}\right) Og\left( \xi _{0},\xi ^{\prime }\right) \label{DIS1} 
\end{equation}%
and solve for $g\left( \xi _{0},\xi ^{\prime }\right)$  as
\begin{equation}
g\left( \xi _{0},\xi ^{\prime }\right) =\frac{1}{\left( 1-\frac{\tilde{\theta%
}}{4\pi }\mathfrak{g}\left( \xi _{0},\xi _{0}\right) O\right) }\mathfrak{g}%
\left( \xi _{0},\xi ^{\prime }\right). \label{DIS2} 
\end{equation}%
Then we substitute the above result in Eq. (\ref{COMGFEQ}) obtaining 
\begin{equation}
g\left( \xi ,\xi ^{\prime }\right) =\mathfrak{g}\left( \xi ,\xi ^{\prime
}\right) +\frac{\tilde{\theta}}{4\pi }\mathfrak{g}\left( \xi ,\xi
_{0}\right) O\frac{1}{\left( 1-\frac{\tilde{\theta}}{4\pi }\mathfrak{g}%
\left( \xi _{0},\xi _{0}\right) O\right) }\mathfrak{g}\left( \xi _{0},\xi
^{\prime }\right), \label{FINSOL}
\end{equation}%
which expresses the reduced GF $\left( g_{\phantom{\mu}\sigma }^{\mu }\right) _{P,T}$ 
in terms of the free GF $\left( \mathfrak{g}_{\phantom{\mu}\sigma }^{\mu }\right)
_{P,T}$. The full GF is reconstructed then from the Eq. (\ref{FULLGF}). In the specific cases considered in Refs.~\citenum{AMU1,AMU2,AMU3} the solutions of Eqs. (\ref{DIS2}) and (\ref{FINSOL}) are explicitly constructed in a step by step fashion to be illustrated in the next section.

\section{The case of a planar interface}
\label{PLS}

The simplest example of the construction previously discussed is when the
interface $\Sigma$ is the plane $z=a$. Here the MEP $\theta (%
\mathbf{x})$ is 
\begin{equation}
\theta (z)=\theta _{1}H(a-z)+\theta _{2}H(z-a),  \label{PWC}
\end{equation}%
where $H(z)$ is the Heaviside function. Then $
\nabla \theta =(\theta _{2}-\theta _{1})\delta (z-a)\hat{\mathbf{e}}_{z}$,
and $\hat{\mathbf{e}}_{z}$ is
the unit vector in the direction $z$. In this way, the dynamical
modifications in Eqs. (\ref{MAXMOD}) arise only at the boundary $z=a$,
which is the only place where the effective sources (\ref{DENEFF}) are
nonzero. That is to say, the $\theta $-vacuum has conducting properties at the
boundary $\Sigma$, even though its bulk behaves as ordinary vacuum.
The general eigenfunctions in Eq. (\ref{EIGENF})  take the form
\begin{equation}
\Psi _{\mathbf{p_{\parallel }}}(x,y)=\frac{1}{\left( 2\pi \right) }e^{i\mathbf{p}%
_{\parallel }\cdot \mathbf{x}_{\parallel}}\;,\;
\end{equation}%
where the index $M$ is now the momentum $\mathbf{p}_{\parallel
}=(p_{x},p_{y})$  parallel to the plane $\Sigma $ and $\mathbf{x}_{\parallel }=(x,y)$.
This adds up to realize the Eq. (\ref{FULLGF}) by introducing the reduced
GF $\left( g_{\phantom{\mu}\nu }^{\mu }\right) _{\mathbf{p,p}^{\prime }}\left(
z,z^{\prime }\right)$ as%
\begin{equation}
G_{\phantom{\mu}\nu }^{\mu }\left( \mathbf{x},\mathbf{x}^{\prime }\right) =4\pi \int
d^{2}\mathbf{p}_{\parallel }\mathbf{\;}d^{2}\mathbf{p}_{\parallel }^{\prime }%
\frac{1}{\left( 2\pi \right) ^{2}}e^{i\mathbf{p}_{\parallel }\cdot \left( 
\mathbf{x}-\mathbf{x}^{\prime }\right) _{\parallel }}\left( g_{\phantom{\mu}\nu }^{\mu
}\right) _{\mathbf{p,p}^{\prime }}\left( z,z^{\prime }\right) 
\label{GFPLANAR}
\end{equation}%
In this case the operator in Eq. (\ref{OP}) is $O_{\phantom{\mu}\ \nu }^{\mu }=\epsilon
_{\phantom{z}\phantom{\mu}\phantom{\alpha}\nu }^{z\mu \alpha }\partial _{\alpha }$  and its matrix
elements  of Eq. (\ref{ME1}) simplify to 
\begin{equation}
\left[ O_{\phantom{\mu}\nu }^{\mu }\right] _{\mathbf{p,p}^{\prime }} =\epsilon
_{\phantom{3}\phantom{\mu}\phantom{\alpha}\nu }^{3\mu \alpha }ip_{\alpha }\delta ^{2}(\mathbf{p}%
_{\parallel }\mathbf{-p}_{\parallel }^{\prime }),
\end{equation}%
where $p^{\alpha }=(0,p_{x},p_{y},0)=(0,\mathbf{p}_{\parallel })$. Since $%
\left[ O_{\phantom{\mu}\nu }^{\mu }\right] _{\mathbf{p,p}^{\prime }}$ is diagonal in
momentum space, Eq.\ (\ref{REDGFFINAL})\ indicates that we can also  take $%
\left( g_{\phantom{\mu}\nu }^{\mu }\right) _{\mathbf{p,p}^{\prime }}\left( z,z^{\prime
}\right) $ to be  diagonal, so that we write%
\begin{equation}
\left( g_{\phantom{\mu}\nu }^{\mu }\right) _{\mathbf{p,p}^{\prime }}\left( z,z^{\prime
}\right) =\delta ^{2}(\mathbf{p}_{\parallel }\mathbf{-p}_{\parallel
}^{\prime })g_{\phantom{\mu}\nu }^{\mu }\left( z,z^{\prime },\mathbf{p}_{\parallel
}\right) .
\end{equation}%
In this way, the final representation for the GF of Eq. (\ref{GFPLANAR}) turns
out to be given in terms of  the Fourier transform in the directions $x, y$ parallel
to the plane $\Sigma $ \cite{CED} 
\begin{equation}
G_{\phantom{\mu}\nu }^{\mu }\left( \mathbf{x},\mathbf{x}^{\prime }\right) =4\pi \int 
\frac{d^{2}\mathbf{p_{\parallel }}}{\left( 2\pi \right) ^{2}}e^{i\mathbf{p}_{\parallel
}\cdot \left( \mathbf{x}-\mathbf{x}^{\prime }\right) _{\parallel }}g_{\phantom{\mu}\nu
}^{\mu }\left( z,z^{\prime },\mathbf{p}_{\parallel }\right) ,
\label{RedGreenDef}
\end{equation}%
as expected.

Due to the antisymmetry of the Levi-Civita symbol, the partial
derivative appearing in the second term of the GF Eq. (\ref{EqsGreenMatrix})
does not introduce derivatives with respect to $z$, but only in the
transverse directions. This allows us to write the full reduced GF equation
as 
\begin{equation}
\left[ \partial ^{2}\eta _{\phantom{\mu}\nu }^{\mu }+i\tilde{\theta}\delta \left(
z-a\right) \epsilon _{\phantom{3}\phantom{\mu}\phantom{\alpha}\nu }^{3\mu \alpha }p_{\alpha }\right]
g_{\phantom{\nu}\sigma }^{\nu }\left( z,z^{\prime },\mathbf{p},\right) =\eta _{\phantom{\mu}
\sigma }^{\mu }\delta \left( z-z^{\prime }\right) ,  \label{RedGreenFunc}
\end{equation}%
where $\partial ^{2}=\mathbf{p}^{2}-\partial _{z}^{2}$, $p^{\alpha }p_{\alpha}=-\mathbf{p_{\parallel}}%
^{2} $ and we denote $|\mathbf{p_{\parallel}}|= p$. 

The solution of Eq. (\ref{RedGreenFunc}) is obtained with the introduction of
a  reduced free GF having the form $\mathfrak{G}_{\phantom{\mu}\nu }^{\mu }\left(
z,z^{\prime }\right) =\mathfrak{g}\left( z,z^{\prime }\right) \eta _{\phantom{\mu}\nu
}^{\mu }$, associated with the operator $\partial ^{2}$ previously defined,
that solves 
\begin{equation}
\partial ^{2}\mathfrak{G}_{\phantom{\mu}\nu }^{\mu }\left( z,z^{\prime }\right) =\eta
_{\phantom{\mu}\nu }^{\mu }\delta \left( z-z^{\prime }\right),  \label{RedGreenFuncA1}
\end{equation}%
plus BC's. In the case of standard BC's at  infinity, the choice is \cite{CED}
\begin{equation}
\mathfrak{g}(z,z^{\prime })=\frac{1}{2p}e^{-p|z-z^{\prime }|}.  \label{RFSGF}
\end{equation}
 Note that Eq. (\ref%
{RedGreenFuncA1}) demands the derivative of $\mathfrak{g}$ to be
discontinuous at $z=z^{\prime} $, \emph{i.e.,} $\partial _{z} \mathfrak{g}
\left( z , z ^{\prime} \right) \big|  _{z= z^{\prime -}} ^{z=z^{\prime +}} =
-1$, together with  the continuity of $\mathfrak{g}$ at $z=z^{\prime}$.

Now we observe that Eq. (\ref{RedGreenFunc}) can be directly integrated by
using the free GF in Eq. (\ref{RedGreenFuncA1}) together with the properties of
the Dirac delta-function, thus reducing the problem to a set of coupled
algebraic equations, 
\begin{equation}
g_{\phantom{\mu}\sigma }^{\mu }\left( z,z^{\prime }\right) =\eta _{\phantom{\mu}\sigma }^{\mu }%
\mathfrak{g}\left( z,z^{\prime }\right) -i\tilde{\theta}\epsilon_{\phantom{3}\phantom{\mu}\phantom{\alpha}
\nu }^{3\mu \alpha }p_{\alpha }\mathfrak{g}\left( z,a\right) g_{\phantom{\nu}\sigma
}^{\nu }\left( a,z^{\prime }\right) .  \label{RedGreenFuncA3}
\end{equation}%
Note that the continuity of $\mathfrak{g}$ at $z=z^{\prime }$ implies the
continuity of $g_{\phantom{\mu}\sigma }^{\mu }$, but the discontinuity of $\partial
_{z}\mathfrak{g}$ at the same point yields 
\begin{equation}
\partial _{z}g_{\phantom{\mu}\sigma }^{\mu }\left( z,z^{\prime }\right) \big|%
_{z=a^{-}}^{z=a^{+}}=-i\tilde{\theta}\epsilon _{\phantom{3}\phantom{\mu}\phantom{\alpha}\nu }^{3\mu \alpha
}p_{\alpha }\partial _{z}\mathfrak{g}\left( z,a\right) \big|%
_{z=a^{-}}^{z=a^{+}}g_{\phantom{\nu}\sigma }^{\nu }\left( a,z^{\prime }\right) =i%
\tilde{\theta}\epsilon _{\phantom{3}\phantom{\mu}\phantom{\alpha}\nu}^{3\mu \alpha }p_{\alpha }g_{\phantom{\nu}
\sigma }^{\nu }\left( a,z^{\prime }\right) ,  \label{BCRedGreenFunc}
\end{equation}%
from which the boundary conditions for the 4-potential in Eq. (\ref{BC4Pot})
are recovered. In this way the solution in  Eq. (\ref{RedGreenFuncA3}) guarantees
that the boundary conditions at the $\theta $-interface are satisfied. 

In this case the formal solution  for Eq.  (\ref{FINSOL})  for $g_{\phantom{\mu} \sigma
}^{\mu }\left( z,z^{\prime }\right) $ can be explicitly obtained in
successive steps. To this end we split Eq. (\ref{RedGreenFuncA3}) into $\mu =0
$ and $\mu =j=1,2,3$ components; 
\begin{eqnarray}
g_{\phantom{0}\sigma }^{0}\left( z,z^{\prime }\right)  &=&\eta _{\phantom{0}\sigma }^{0}%
\mathfrak{g}\left( z,z^{\prime }\right) -i\tilde{\theta}\epsilon _{\phantom{3}\phantom{0}\phantom{i}
j}^{30i}p_{i}\mathfrak{g}\left( z,a\right) g_{\phantom{j}\sigma }^{j}\left(
a,z^{\prime }\right) ,  \label{g0S} \\
g_{\phantom{j}\sigma }^{j}\left( z,z^{\prime }\right)  &=&\eta _{\phantom{j}\sigma }^{j}%
\mathfrak{g}\left( z,z^{\prime }\right) -i\tilde{\theta}\epsilon _{\phantom{3}\phantom{j}\phantom{i}
0}^{3ji}p_{i}\mathfrak{g}\left(z,a\right) g_{\phantom{0}\sigma }^{0}\left(
a,z^{\prime }\right) .  \label{gjS}
\end{eqnarray}%
Now we set $z=a$ in Eq. (\ref{gjS}) and then substitute into Eq. (\ref{g0S})
yielding 
\begin{equation}
g_{\phantom{0}\sigma }^{0}\left( z,z^{\prime }\right) =\eta _{\phantom{0}\sigma }^{0}\mathfrak{%
g}\left( z,z^{\prime }\right) -i\tilde{\theta}\epsilon _{\phantom{3}\phantom{0}\phantom{i}
j}^{30i}p_{i}\eta _{\phantom{j}\sigma }^{j}\mathfrak{g}\left( z,a\right) \mathfrak{g}%
\left( a,z^{\prime }\right) -\tilde{\theta}^{2}p^{2}\mathfrak{g}\left(
z,a\right) \mathfrak{g}\left( a,a\right) g_{\phantom{0}\sigma }^{0}\left( a,z^{\prime
}\right) ,  \label{g0S-2}
\end{equation}%
where we use the result $\epsilon_{\phantom{3}\phantom{0}\phantom{i} j}^{30i}\epsilon_{\phantom{3}\phantom{j}\phantom{k}
0}^{3jk}\,p_{k}p_{i}=p^{2}$. Solving for $g_{\phantom{0}\sigma }^{0}\left( a,z^{\prime
}\right) $ by setting $z=a$ in Eq. (\ref{g0S-2}) and inserting the result
back in  Eq. (\ref{g0S-2}), we obtain 
\begin{equation}
g_{\phantom{0}\sigma }^{0}\left( z,z^{\prime }\right) =\eta _{\phantom{0}\sigma }^{0}\left[ 
\mathfrak{g}\left( z,z^{\prime }\right) +\tilde{\theta}p^{2}\mathfrak{g}%
\left( a,a\right) A\left( z,z^{\prime }\right) \right] +i\epsilon_{\phantom{3}\phantom{0}\phantom{i}
\sigma }^{30i}p_{i}A\left( z,z^{\prime }\right) ,  \label{g0S-3}
\end{equation}%
where 
\begin{equation}
A\left( z,z^{\prime }\right) =-\tilde{\theta}\frac{\mathfrak{g}\left(
z,a\right) \mathfrak{g}\left( a,z^{\prime }\right) }{1+p^{2}\tilde{\theta}%
^{2}\mathfrak{g}^{2}\left( a,a\right) }.  \label{A(Z,Z)}
\end{equation}%
The remaining components can be obtained by substituting $g_{\phantom{0}\sigma
}^{0}\left( a,z^{\prime }\right) $ in Eq. (\ref{gjS}). The result is 
\begin{equation}
g_{\phantom{j}\sigma }^{j}\left( z,z^{\prime }\right) =\eta _{\phantom{j}\sigma }^{j}\mathfrak{%
g}\left( z,z^{\prime }\right) +i\epsilon_{\phantom{3}\phantom{j}\phantom{k}0}^{3jk}p_{k}\left[ \eta
_{\phantom{0}\sigma }^{0}-i\tilde{\theta}\epsilon_{\phantom{3}\phantom{0}\phantom{i}\sigma }^{30i}p_{i}%
\mathfrak{g}\left( a,a\right) \right] A\left( z,z^{\prime }\right) .
\label{gjS-2}
\end{equation}%
Equations (\ref{g0S-3}) and (\ref{gjS-2}) allow us to write the general
solution as 
\begin{equation}
g_{\phantom{\mu}\nu }^{\mu }\left( z,z^{\prime }\right) =\eta _{\phantom{\mu}\nu }^{\mu }%
\mathfrak{g}\left( z,z^{\prime }\right) +A\left( z,z^{\prime }\right)
\left\{ \tilde{\theta}\mathfrak{g}\left( a,a\right) \left[ p^{\mu }p_{\nu
}+\left( \eta_{\phantom{\mu}\nu }^{\mu }+n^{\mu }n_{\nu }\right) p^{2}\right]
+i\epsilon_{\phantom{\mu} \nu }^{\mu \phantom{\nu} \alpha 3}p_{\alpha }\right\} ,
\label{GenSolGreenPlaneConf}
\end{equation}%
where $n_{\mu }=\left( 0,0,0,1\right) $ is the normal to $\Sigma $%.

The reciprocity between the position of the unit charge and the position at
which the GF is evaluated $G_{\mu \nu }(\mathbf{x},\mathbf{x} ^{\prime }) =
G_{\nu \mu }(\mathbf{x} ^{\prime},\mathbf{x})$ is one of its most remarkable
properties of the GF. From Eq. (\ref{RedGreenDef}) this condition demands 
\begin{equation}
g_{\mu \nu }(z,z^{\prime }, \mathbf{p})=g_{\nu \mu }(z^{\prime },z,- \mathbf{%
p}),
\end{equation}%
which we verify directly from Eq. (\ref{GenSolGreenPlaneConf}). The symmetry 
$g_{\mu \nu }\left( z,z^{\prime }\right) =g_{\nu \mu }^{\ast }\left(
z,z^{\prime }\right) =g_{\mu \nu }^{\dagger }\left( z,z^{\prime }\right) $
is also manifest.

The various components of the static GF matrix in coordinate representation
are obtained by computing the Fourier transform defined in Eq. (\ref%
{RedGreenDef}), with the reduced GF given by Eq. (\ref{RFSGF}). The details
are presented in Ref.~\citenum{AMU1}. The final results are 
\begin{eqnarray}
G_{\phantom{0}0}^{0}\left( \mathbf{x},\mathbf{x}^{\prime }\right)  &=&\frac{1}{|%
\mathbf{x}-\mathbf{x}^{\prime }|}-\frac{\tilde{\theta}^{2}}{4+\tilde{\theta}%
^{2}}\frac{1}{\sqrt{R^{2}+Z^{2}}},  \label{G00} \\
G_{\phantom{0}i}^{0}\left( \mathbf{x},\mathbf{x}^{\prime }\right)  &=&-\frac{2\tilde{%
\theta}}{4+\tilde{\theta}^{2}}\frac{\epsilon _{0ij3}R^{j}}{R^{2}}\left( 1-%
\frac{Z}{\sqrt{R^{2}+Z^{2}}}\right) ,  \label{G0i} \\
G_{\phantom{i}j}^{i}\left( \mathbf{x},\mathbf{x}^{\prime }\right)  &=&\eta
_{\phantom{i}j}^{i}G_{\;0}^{0}\left( \mathbf{x},\mathbf{x}^{\prime }\right) -\frac{i}{%
2}\frac{\tilde{\theta}^{2}}{4+\tilde{\theta}^{2}}\partial ^{i}K_{j}\left( 
\mathbf{x},\mathbf{x}^{\prime }\right) ,  \label{Gij}
\end{eqnarray}%
where $Z=|z-a|+|z^{\prime }-a|$, $R^{j}=\left( \mathbf{x-x}^{\prime }\right)
_{\parallel }^{j}=\left( x-x^{\prime },y-y^{\prime }\right) $, $R=|\left( 
\mathbf{x-x}^{\prime }\right) _{\parallel }|\;$and%
\begin{equation}
K^{j}\left( \mathbf{x},\mathbf{x}^{\prime }\right) =2i\frac{\sqrt{R^{2}+Z^{2}%
}-Z}{R^{2}}R^{j}.
\end{equation}%
Finally, we observe that Eqs. (\ref{G00}-\ref{Gij}) contain all the required
elements of the GF matrix, according to the choices of $z$ and $z^{\prime }$
in the function $Z$.

Similar results for the cases of spherical and cylindrical interfaces incorporating also 
piecewise continuous ponderable media have been reported in Refs.~\citenum{AMU2,AMU3}.

\section{The Casimir effect }
\label{CASE}

The Casimir effect
(CE)
\cite{Casimir} 
is one of the most
remarkable consequences of the nonzero
vacuum energy predicted by quantum
field theory which has been confirmed by
experiments \cite{Bressi}. 
 In general, the CE can be defined as the stress (force
per unit  area) on bounding surfaces
when a quantum field is confined in a
finite volume of space. The boundaries
can be  material media, interfaces between
two phases of  vacuum, or topologies
of space. For a review  see, for example, 
Refs.~\citenum{Milton,Bordag}. 

The experimental accessibility to 
micrometer-size physics together with 
the recent discovery of three dimensional TIs
\cite{Liang} provides an additional arena where the CE can be studied.
In  the scattering approach to the Casimir effect, \textit{i.e.} using the Fresnel coefficients for
the reflection matrices at the interfaces of the TIs, the Casimir force between
TIs was computed in Ref.~\citenum{Grushin}. The authors found 
the most notable feature that, due to the magnetoelectric effect, which now has a topological origin, the
strength and sign of the Casimir stress between two planar TIs can be tuned. 

When the surface of the TI is included in the description,
$\theta$-ED is a fair description of both the
bulk and the surface only when a
time reversal symmetry breaking
perturbation is induced on the surface to
gap the surface states, thereby
converting it into a full insulator. In this
situation, which we consider
here, the MEP $\theta$ can be
shown to be quantized in odd integer
values of $\pi$:
$\theta = (2n+1) \pi$, where $n \in \mathbb{Z}$ is
determined by the nature
of the time reversal symmetry breaking
perturbation, which could be controlled 
experimentally by covering the TI with a
thin magnetic layer \cite{Grushin}. For a review of the effective $\theta$-ED
describing the electromagnetic response of  TI's see Refs.~\citenum{TI, TI1} for example.

The Casimir system
we consider is formed by
two perfectly reflecting planar surfaces
(labeled $P_{1}$ and
$P_{2}$) separated by a
distance $L$, with 
 a non-trivial TI placed between
them, but perfectly joined to the
plate $P_{2}$, as shown in Fig. \ref{fig1}.  The
surface $\Sigma $ of the TI, located at $z=a$, is assumed to be covered by a
thin magnetic layer which breaks time reversal
symmetry there. We  calculate  the Casimir stress
restricting ourselves only to the contribution of the MEP which now has a topological origin, \textit{i.e.} we set $\varepsilon=\mu=1$. 
We
follow an approach similar
to that in Ref.~\citenum{Brown} which starts from the calculation of  the appropriate 
GF, to subsequently compute
 the renormalized
vacuum stress-energy tensor in the region between the plates yielding finally the Casimir stress that the plates exert on the surface $\Sigma$ of the TI. We  also consider  the limit
where the plate $P_{2}$ is sent to
infinity ($L \rightarrow \infty $)  to obtain the Casimir stress
between a conducting plate and a
non-trivial semi-infinite TI. 

The  BCs for the perfectly reflecting metallic plates $P_1$
and $P_2$ are the standard ones $n _{\mu} \tilde{F}
^{\mu \nu} \vert _{P_{1,2}}
= 0$, where $ n_{\mu} = (0,0,0,1)$.
 The effects of the MEP are 
incorporated by choosing 
\begin{equation}
\theta (z) = \theta H (z - a) H (L - z), \quad \tilde{\theta} = - \alpha \theta / \pi.
 \label{thetaCAS}
\end{equation}

\begin{figure}[tbp]
\begin{center}
\includegraphics[width=2.5in]{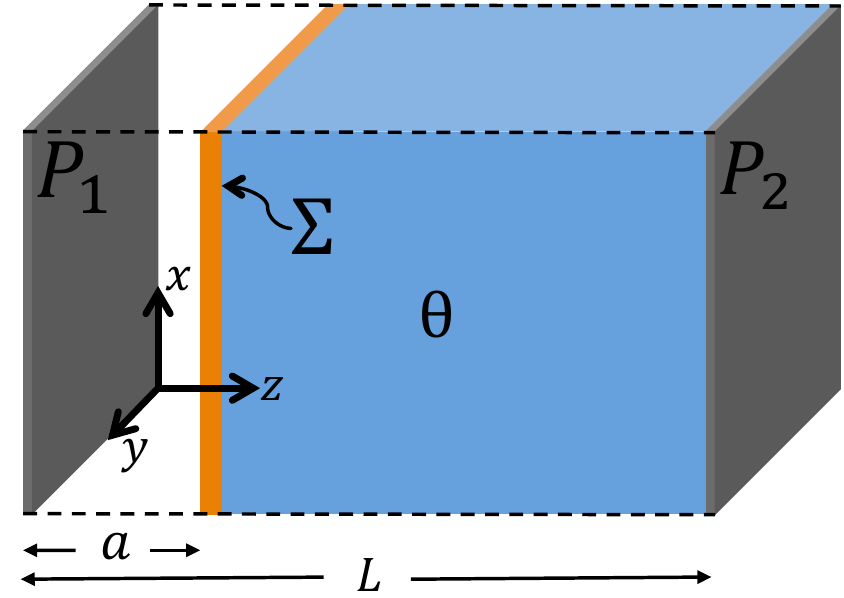}
\end{center}
\caption{{\protect\small Schematic of the Casimir effect in $\theta$-ED}.}
\label{fig1}
\end{figure}

Assuming 
the absence of free
sources on $\Sigma$,
the required equation for the  GF matrix is given by Eq. 
 (\ref{EqsGreenMatrix}),
together with the BCs arising from  Eq. (\ref{BC4Pot}). The calculation proceeds along
the same lines discussed in section \ref{PLS} for the static case, but keeping the time dependence now.
Making explicit the coordinate choice in the
transverse $x$ and $y$ directions we
can write
\begin{equation}
G _{\phantom{\mu} \nu }^{\mu }\left( x , x ^{\prime }\right) =
4 \pi \int \frac{d^{2}\mathbf{p_{\parallel}}}{\left( 2\pi \right)^{2}}
e^{i\mathbf{p_{\parallel}} \cdot \mathbf{x_{\parallel}}} \int \frac{d \omega}{2 \pi}
e^{-i \omega ( t - t ^{\prime} )}
g_{\phantom{\mu} \nu }^{\mu}\left( z,z^{\prime }\right) ,  \label{RedGreenDefCAS}
\end{equation}%
where we have omitted 
the dependence of the reduced GF
$g_{\phantom{\mu} \nu }^{\mu }$ on $\omega$ and $
\mathbf{p_{\parallel}}$. In the Lorenz gauge the equation for the
reduced GF $g_{\phantom{\nu} \sigma }^{\nu }
( z,z^{\prime })$ is
\begin{equation}
\left[ \eta _{\phantom{\mu}\nu }^{\mu } \partial ^{2} + i
\tilde{\theta}\delta \left( z-a \right)
\epsilon _{\phantom{3\mu \alpha} \nu }^{3 \mu \alpha}
p_{\alpha }\right] g_{\phantom{\nu} \sigma }^{\nu }
\left( z,z^{\prime }\right) =
\eta _{\phantom{\mu} \sigma }^{\mu}
\delta \left( z-z^{\prime }\right) ,  \label{RedGreenFuncCAS}
\end{equation}
where now $\partial ^{2}= \mathbf{p_{\parallel}}
^{2} - \omega ^{2} - \partial _{z}^{2}$
and $p^{\alpha}=\left( \omega ,
\mathbf{p_{\parallel}} , 0 \right) $. The boundary term
(at $z=L$), missing
in Eq. (\ref{RedGreenFuncCAS}),
identically vanishes in the distributional sense,  due to the BCs on the
plate $P_{2}$. In this way,
Eq. (\ref{RedGreenFuncCAS}) implies that
the only topologically magnetoelectric effect  present in our Casimir
system is the one produced at $\Sigma$.

Here  the free GF  we  use to integrate Eq. (\ref{RedGreenFuncCAS})
is the reduced GF for two parallel
conducting surfaces placed at
$z=0$ and $z=L$, which is the solution of
$\partial ^{2} \mathfrak{g} \left( z ,
z^{\prime} \right) = \delta \left( z - z
^{\prime} \right)$ satisfying the BCs
$\mathfrak{g} \left(
0 , z^{\prime} \right) = \mathfrak{g}
\left( L , z^{\prime} \right) = 0$, namely \cite{CED}
\begin{equation}
\mathfrak{g}_c \left( z , z ^{\prime} \right) = \frac{\sin
\left[ p z _{<} \right] \sin \left[ p \left( L - z _{>} \right) \right]}%
{p \sin \left[ p L \right]} , \label{RGF-plates}
\end{equation}
where $z _{>}$ ($z _{<}$) is the greater
(lesser) of $z$ and $z ^{\prime}$, and $p
= \sqrt{\omega ^{2} - \textbf{p} ^{2}}$.
Now the problem is reduced to a set of
coupled algebraic equations,
\begin{equation}
g ^{\mu} _{\phantom{\mu} \sigma} \left( z , z ^{\prime} \right) =
\eta ^{\mu} _{\phantom{\mu} \sigma} \mathfrak{g}_c
\left( z , z ^{\prime} \right) - i \tilde{\theta}
\epsilon ^{3 \mu \alpha} _{\phantom{3 \mu \alpha} \nu }
p_{\alpha } \mathfrak{g}_c \left( z , a \right)
g ^{\nu} _{\phantom{\nu} \sigma} \left( a , z ^{\prime} \right) .  \label{A3}
\end{equation}

We write the general solution to Eq. (\ref{A3})  as the sum of two terms
\begin{equation}
g^{\mu} _{\phantom{\mu}\nu } \left( z , z ^{\prime} \right) =
\eta ^{\mu } _{\phantom{\mu} \nu } \mathfrak{g}_c
\left( z , z ^{\prime } \right) + {g}^{\mu}_{C \nu }
\left( z , z^{\prime} \right).\label{DESREDGF}
\end{equation} 
The first term provides the propagation in 
the absence of the TI between the parallel plates.
The second, 
to be called  the reduced $\theta$-GF, which 
can be shown to be
\begin{eqnarray}
g^{\mu}_{C \nu} \left( z , z^{\prime} \right)=
\tilde{\theta} \mathfrak{g}_c \left( a , a \right) \left[ p^{\mu} p_{\nu}-
\left( \eta ^{\mu} _{\phantom{\mu} \nu } + n ^{\mu} n _{\nu} \right)
p ^{2} \right] A_c \left( z , z^{\prime} \right) + i \,
\epsilon ^{\mu \phantom{\nu} \alpha 3} _{\phantom{\mu} \nu }
p_{\alpha } A_c \left( z , z^{\prime} \right), \nonumber \\ \label{theta-GF}
\end{eqnarray}
encodes the  magnetoelectric effect  due to the
topological MEP $\theta$. Here
\begin{equation}
A_c \left( z,z^{\prime }\right) = - \tilde{\theta} \frac{\mathfrak{g}_c
\left( z , a \right) \mathfrak{g}_c\left( a,z^{\prime }\right) }%
{1- p^{2} \tilde{\theta}^{2}\mathfrak{g}_c^{2}\left( a,a\right) }, 
\label{A(Z,Z)CAS}
\end{equation}
has the same form as the previous Eq. (\ref{A(Z,Z)}) with   $\mathfrak{g}
\left(z , z' \right) \rightarrow \mathfrak{g}_c
\left(z , z' \right) $. 
In the static limit ($\omega = 0$), our result in Eq. (\ref{theta-GF})
reduces to the one reported in Ref.~\citenum{AMU1}. As the Eq. (\ref{DESREDGF}) suggests,  the full GF matrix  $G ^{\mu} _{\phantom{\mu} \nu} \left( x , x ^{\prime} \right)$  can also be
written as the sum of two terms,
$G ^{\mu} _{\phantom{\mu} \nu} \left( x , x ^{\prime} \right) =
\eta ^{\mu} _{\phantom{\mu}\nu}
\mathcal{G} \left( x , x ^{\prime} \right) +
G^{\mu}_{C \nu} \left( x , x ^{\prime} \right)$, each one arising from the respective  term in the Eq. (\ref{DESREDGF}). We call $G^{\mu}_{C \nu} \left( x , x ^{\prime} \right)$ the $\theta$-GF.

Since the MEP modifies the behavior of the fields only at the interface,
we expect that stress energy tensor (SET) in the bulk retains its original Mawxwell's form.
In fact, in Ref.~\citenum{AMU2} we explicitly computed  
the SET and verified that
\begin{equation}
T ^{\mu \nu} = \frac{1}{4 \pi} \left( - F ^{\mu \lambda}
F ^{\nu} _{\phantom{\nu}\lambda} + \frac{1}{4}
\eta^{\mu \nu} F _{\alpha \beta} F ^{\alpha \beta} \right) .
\label{Stress-Energy}
\end{equation}
Clearly this tensor is traceless and its divergence
is
\begin{equation}
\partial _{\mu} T ^{\mu \nu} = - F ^{\nu} _{\phantom{\nu} \lambda}
j ^{\lambda} - ( \tilde{\theta} / 4 \pi ) \delta \left( \Sigma \right)
n _{\mu} F ^{\nu} _{\phantom{\nu} \lambda}
\tilde{F} ^{\mu \lambda} . \label{DivST}
\end{equation}
As expected, $T^{\mu\nu}$
it is not conserved at $\Sigma$
because  the MEP  induces effective charge
and current densities there.

Now we address the calculation of the
vacuum expectation
value of the  SET,  to
which we will refer simply as the vacuum
stress (VS).
The local approach to compute the VS
was initiated by Brown and Maclay who
calculated the renormalized stress tensor  by
means of GF techniques
\cite{Brown,Schwinger1}. 
Using the standard point splitting procedure and  taking the vacuum expectation value of the SET in
(\ref{Stress-Energy}) we find
\begin{eqnarray}
\left<  T ^{\mu \nu} \right> = \frac{i}{4 \pi} \lim _{x \rightarrow x ^{\prime}}
\Big[ - \partial ^{\mu}  \partial ^{\prime \nu}
G ^{\lambda} _{\phantom{\lambda} \lambda} + \partial ^{\mu}
\partial _{\lambda} ^{\prime} G ^{\lambda \nu} +
\partial ^{\lambda} \partial ^{\prime \nu} G ^{\mu} _{\phantom{\mu} \lambda}
\nonumber \\
-\,  \partial ^{\prime \lambda} \partial _{\lambda} G ^{\mu \nu} + \frac{1}{2}
\eta ^{\mu \nu} \left( \partial ^{\alpha} \partial _{\alpha} ^{\prime}
G ^{\lambda} _{\phantom{\lambda} \lambda} -
\partial ^{\alpha} \partial _{\beta} ^{\prime}
G ^{\beta} _{\phantom{\beta} \alpha} \right) \Big] ,
\label{VacuumStress}
\end{eqnarray}
where we have omitted the
dependence of $G^{\mu \nu}$ on
$x$ and $x ^{\prime}$. This result can be further simplified
as follows. Since the GF is written as the sum of two terms, the VS can
also be written in the same way, \textit{i.e.,}
\begin{equation}
\left< T ^{\mu \nu} \right> = \left< t ^{\mu \nu} \right> + %
\left<T_C^{\mu \nu}  \right>.
\end{equation}
The first term,
\begin{equation}
\left< t ^{\mu \nu} \right> = \frac{1}{4 \pi i} \lim _{x \rightarrow x ^{\prime}}
\left( 2 \partial ^{\mu} \partial ^{\prime \nu} - \frac{1}{2} \eta ^{\mu \nu}
\partial ^{\lambda} \partial _{\lambda} ^{\prime} \right) \mathcal{G}
\left( x , x ^{\prime} \right) , \label{SE-free}
\end{equation}
is the VS in the absence of the
TI. In obtaining Eq.
(\ref{SE-free}) we use that the
GF is diagonal when the TI is absent, \textit{i.e.} it is equal to
$\eta^{\mu} _{\phantom{\mu} \nu} \mathcal{G}
\left( x , x^{\prime} \right)$. The second term $\left<T_C^{\mu \nu}  \right>$, to
which we will refer as the
$\theta$ vacuum stress ($\theta$-VS), can be
simplified since the $\theta$-GF satisfies
the Lorenz gauge condition $\partial _{\mu} G_C^{\mu \nu} = 0 $.

With the previous results
the $\theta$-VS can be written as
\begin{equation}
\left< T_C^{\mu \nu} \right> = \frac{1}{4 \pi i}
\lim _{x \rightarrow x ^{\prime}}  \left[ \partial ^{\mu}
\partial ^{\prime \nu} G _{C} + \partial ^{\prime \lambda}
\partial _{\lambda} \left( G_C^{\mu \nu}  - \frac{1}{2}
\eta ^{\mu \nu} G _{C} \right) \right] , \label{VacuumStress2}
\end{equation}
where $G_C = G^{\mu}_{C \mu}$ is the trace of the $\theta$-GF.
This result exhibits the
vanishing of the trace at quantum level,
\textit{i.e.,} $\eta _{\mu
\nu} \left<T_C^{\mu \nu}\right> = 0
$.

 Next we consider  the problem of calculating the
renormalized VS $\left< T^{\mu \nu} \right>
_{\mathrm{ren}}$. We proceed along
the lines of Refs.~\citenum{Brown,Candelas}.
From Eq. (\ref{VacuumStress2}),
together with the symmetry of the
problem we find that  the 
 $\theta$-VS  can be written as
\begin{eqnarray}
\left< T_C^{\mu \nu}  \right> = i \tilde{\theta} \int
\frac{d ^{2} \mathbf{p_{\parallel}}}{(2 \pi) ^{2}} \int \frac{d \omega}{2 \pi}
\left( p ^{\mu} p ^{\nu} + n ^{\mu} n ^{\nu} p ^{2} \right)
\mathfrak{g}_c \left( a , a \right)
\lim _{z \rightarrow z ^{\prime}}  \left( p ^{2} +
\partial ^{\prime} _{z} \partial _{z} \right)
A_c \left( z , z ^{\prime} \right). \nonumber \\  \label{VS2}
\end{eqnarray}
 In deriving
this result we used the Fourier
representation of the GF in Eq. 
(\ref{RedGreenDef}) together with the
solution for the reduced $\theta$-GF
given by Eq. (\ref{theta-GF}). From Eq. (\ref{VS2}) we calculate the
renormalized  $\theta$-VS, which is
given by $\left< T_C^{\mu \nu}
 \right> _{\mathrm{ren}} = \left<T_C^{\mu \nu}
 \right> - \left<T_C
^{\mu \nu} \right>
_{\mathrm{vac}}$, where the  first (second)
term is the $\theta$-VS in the
presence (absence) of the plates  \cite{Candelas}.
When the plates are absent, the reduced
GF we have to use to compute the $
\theta$-VS in the region $[0,L]$ is that
of the free-vacuum $\mathfrak{g}_0(z,z')=(i/2p){\rm exp}(ip|z-z'|)$, from
which we find that $ \lim
_{z \rightarrow z ^{\prime}} \partial _{z}
\partial _{z} ^{\prime} A _{0} \left( z , z
^{\prime} \right) = - p^{2} \lim _{z
\rightarrow z ^{\prime}} A_{0} \left( z ,
z ^{\prime} \right) $, thus implying that
the integrand in Eq. (\ref{VS2}) vanishes.  The function
$A _{0}$ is given by  Eq.
(\ref{A(Z,Z)}) using  the free-vacuum
reduced GF $\mathfrak{g}_0(z,z^{\prime})$.
Therefore we conclude that $ \left< T
^{\mu \nu}_C \right> _{vac} =
0$.

Next we compute $\left< T ^{\mu \nu}_{C} \right> _{\mathrm{ren}}=\left< T ^{\mu \nu} _{C} \right>$ starting from Eq. (\ref{VS2}). From the
symmetry of the problem,
the components of the stress
along the plates, $\left< T^{11}
_{C} \right>$ and $
\left< T ^{22} _{C} \right>
 $, are equal. In addition, from the mathematical
structure of Eq. (\ref{VS2}) we find the
relation $\left< T ^{00}
_{C} \right> = - \left< T^{11} _{C}
\right> $.
These results, together with the traceless
nature of the SET, allow us to write the
renormalized $\theta$-VS in
the form
\begin{equation}
\left< T^{\mu \nu} _{C} \right> _{\mathrm{ren}} = \left( \eta ^{\mu \nu} + 4 n ^{\mu} n ^{\nu} \right) \tau ( \theta , z)  \; , \label{VS3}
\end{equation}
where
\begin{eqnarray}
\tau( \theta , z) &=& i \tilde{\theta} \int \frac{d ^{2} \mathbf{p_{\parallel}}}{(2 \pi) ^{2}} \int \frac{d \omega}{2 \pi} \omega ^{2} \mathfrak{g}_c \left( a , a \right) 
 \lim _{z \rightarrow z ^{\prime}}  \left( p ^{2} + \partial ^{\prime} _{z} \partial _{z} \right) A_c \left( z , z ^{\prime} \right).  \label{f-func}
\end{eqnarray}
Our $\theta$-VS exhibits the
same tensor structure as  the result
obtained by Brown and Maclay \cite{Brown}, 
but now a  $z$-dependent VS arises since
the SET is not conserved at $\Sigma$.
Using Eq.
(\ref{RGF-plates}) we compute the
limit of the integrand {in Eq. (\ref{f-func})}
obtaining
\begin{eqnarray}
\lim _{z \rightarrow z ^{\prime}} \left( p ^{2} + \partial _{z} \partial _{z} ^{\prime} \right) P \left( z , z ^{\prime} \right) = - \frac{\tilde{\theta}}{1 - \tilde{\theta} ^{2} p ^{2} \mathfrak{g}_c^{2} \left( a , a \right)} \times \nonumber \\  \left\lbrace \frac{\sin ^{2} \left[p \left( L - a \right) \right]}{\sin ^{2} \left[p L \right]} H \left( a - z \right) + \frac{\sin ^{2} \left[ p a \right]}{\sin ^{2} \left[p L \right]} H \left( z - a \right) \right\rbrace. \label{Derivatives}
\end{eqnarray}
{To evaluate the integral in Eq.~(24) 
we first write the momentum element as $d ^{2} \mathbf{p_{\parallel}} = \vert \mathbf{p_{\parallel}} \vert d \vert \mathbf{p_{\parallel}} \vert d \vartheta$  and
integrate $\vartheta$. Next, we
perform a Wick rotation such that $\omega \rightarrow i \zeta $, then replace $\zeta$ and $\vert \mathbf{p_{\parallel}} \vert$ by plane polar coordinates $\zeta = \xi \cos \varphi$, $\vert \mathbf{p_{\parallel}} \vert= \xi \sin \varphi$ and finally integrate
$\varphi$. } The renormalized $\theta$-VS in Eq.~(\ref{VS3})
then becomes
\begin{eqnarray}
\left< T^{\mu \nu}_{C} \right>_{\mathrm{ren}}=- \frac{\pi ^{2}}{720 L ^{4}} \left( \eta ^{\mu \nu} + 4 n ^{\mu} n ^{\nu} \right)  \left[ u ( {\theta} , \chi ) H \left( a - z \right) +  u ( {\theta} , 1 - \chi ) H \left( z - a \right) \right], \nonumber \\ \label{VS4}
\end{eqnarray}
where
\begin{equation}
u( {\theta} , \chi ) = \frac{120}{\pi ^{4}} \int _{0} ^{\infty}  \frac{{\tilde{\theta}}^{2} \xi ^{3} \mbox{\small sh} \left[ \xi \chi \right] \mbox{\small sh}^{3} \left[ \xi \left( 1 - \chi \right) \right] \mbox{\small sh}^{-3} \left[ \xi \right]}{1 + {\tilde{\theta}}^{2} \mbox{\small sh} ^{2} \left[ \xi \chi \right] \mbox{\small sh}^{2} \left[ \xi \left( 1 - \chi \right) \right] \mbox{\small sh}^{-2} \left[ \xi \right] } d \xi , \label{funcionU}
\end{equation}
with $\mbox{\small sh}(x) = \sinh (x)$ 
and $\chi = a / L$ with $0 < \chi
< 1$. 
Physically, we interpret the function
$u( {\theta}, \chi)$ as the ratio
between the renormalized
$\theta$-energy density in the vacuum
region $[0 , a )$ and that of the
renormalized energy density in the
absence of the TI. 
The function $u ({\theta} , 1 - \chi
)$ has an analogous interpretation for the
bulk region of the TI $(a , L]$. 
This shows that the energy
density is constant in the bulk regions,
however a simple
discontinuity arises at $\Sigma$, \textit{i.e.,}
$\partial _{z} \left< T ^{00}
_{C} \right> _{\mathrm{ren}} \propto \delta (\Sigma)$.
The Casimir energy  ${\cal E}={\cal E}_L+{\cal E}_{\theta}$ is defined as the
energy per unit  area stored in the
electromagnetic field between the plates.
To obtain it we must integrate the contribution from 
the
$\theta$-energy  density 
\begin{equation}
\mathcal{E}_{\theta}=\int _{0} ^{L} dz \left< T^{00} _{C} \right> _{\mathrm{ren}} = {\mathcal{E}}_L \left[ \chi  u ( {\theta} , \chi ) + (1 - \chi)  u ( {\theta} , 1 - \chi ) \right]. \label{CasEnergy}
\end{equation}
 The first term
corresponds to the energy
stored in the electromagnetic field
between $P _{1}$
and $ \Sigma$, while the second term is
the energy stored in the bulk of the TI.
The ratio $
\mathcal{E} _{\theta} / {\mathcal{E}}_L$ as
a function of $\chi$ for different values of $
\theta$ (appropriate for TIs {\cite{Grushin}}) is 
plotted 
in Fig. \ref{FigCasEnergy} \cite{EPL2}. Let us recall  that $\mathcal{E}_L= -\pi^2/(720 L^3)$ is the Casimir energy
in the absence of the TI.

\begin{figure}[tbp]
\begin{center}
\includegraphics[width=3.0in]{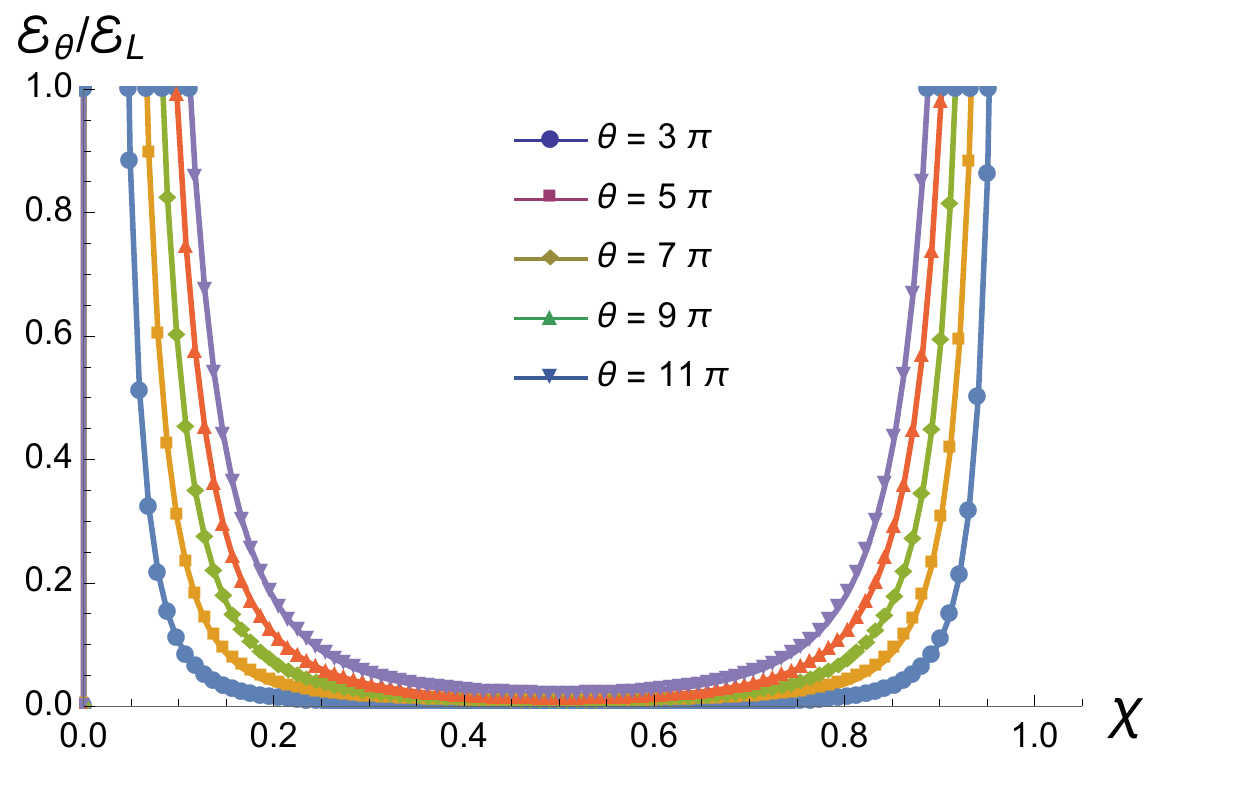}
\end{center}
\caption{{\protect\small The ratio $
\mathcal{E}_{\theta} / \mathcal{E}_{L}$ as
a function of the dimensionless distance $
\chi = a / L$, for different values of $
\theta$. }}
\label{FigCasEnergy}
\end{figure}

\begin{figure}[tbp]
\begin{center}
\includegraphics[width=3.0in]{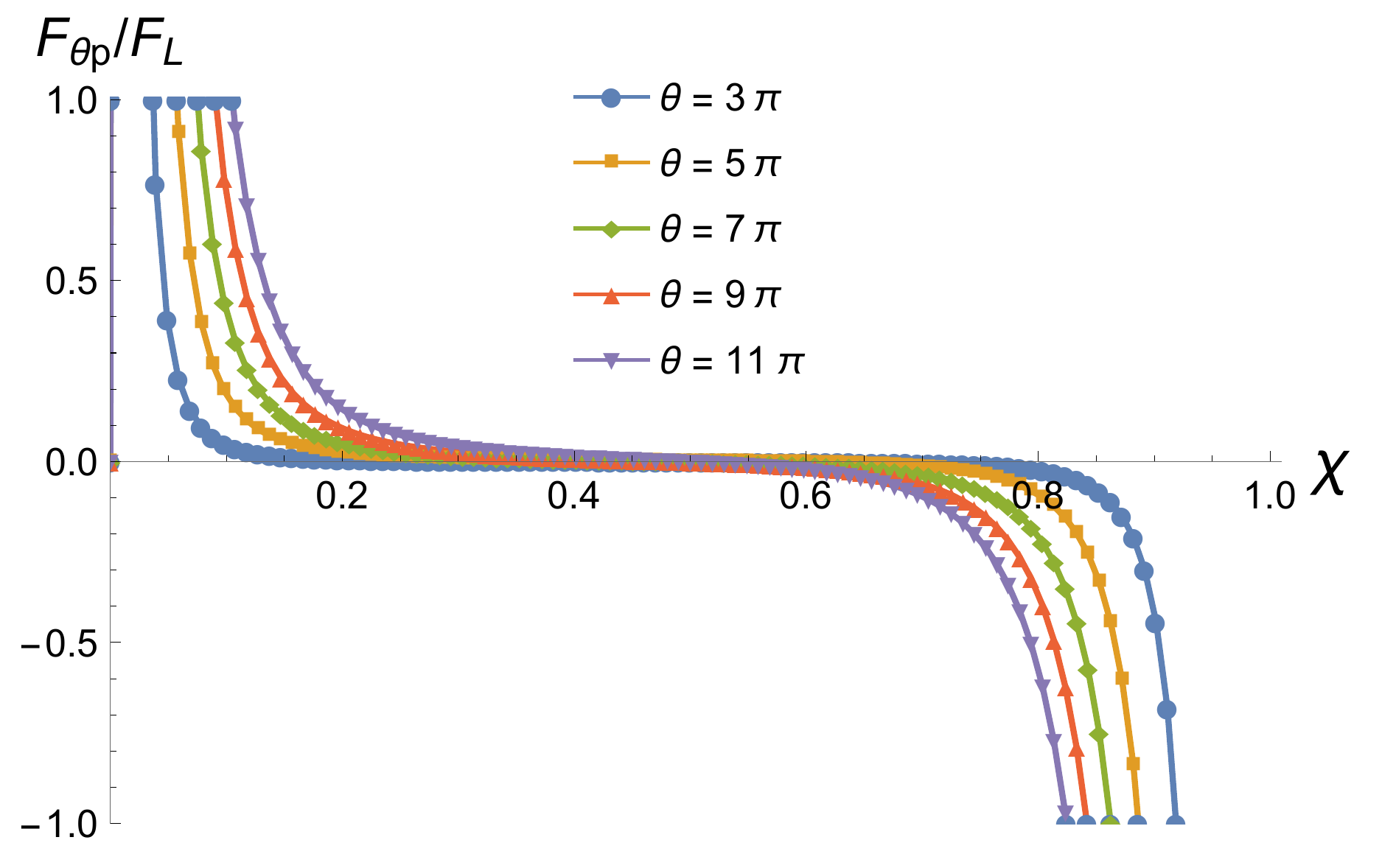}
\end{center}
\caption{{\protect\small  The
Casimir stress on the $\theta$-piston  in units of $F_L$
as a function of $\chi= a / L$, for different
values of $\theta$}.}
\label{ForcePiston}
\end{figure}
The setup known in the literature as the Casimir piston consists of a
rectangular box of length $L$ divided by a
movable mirror (piston) at a distance $a$
from one of the plates \cite{Cavalcanti}.
 The net result is
that the Casimir energy  in each region generates a force
on the piston pulling it towards the
nearest end of the box. Here we have considered
a similar setup, which we call the
$\theta$-piston, in which the piston is 
the  TI. The 
Casimir stress  acting upon
$\Sigma$ can be obtained as 
$F_{\theta p} = - d
\mathcal{E} _{\theta} / da$. The result is
\begin{equation}
\frac{F _{\theta p} }{F_L} = - \frac{1}{3} \frac{d}{d \chi} \left[ \chi  u ( \theta , \chi ) + (1 - \chi)  u ( \theta , 1 - \chi ) \right] ,
\end{equation}
where $F_L--\pi^2/(240 L^4) $ is
the Casimir stress between the two
perfectly reflecting plates in the absence
of the TI.
Figure \ref{ForcePiston} \cite{EPL2} shows the
Casimir stress on $\Sigma$ in units of $F_L$
as a function of $\chi$ for different
values of $\theta$. We observe that this force
pulls the boundary $\Sigma$ towards the closer of the two
fixed walls $P_1$ or $P_2$, similarly to the conclusion in Ref.~\citenum{Cavalcanti}.

Now let us consider the limit where
the plate $P_{2}$ is sent 
to infinity, 
\textit{i.e.,} $L
\rightarrow \infty$. This
configuration corresponds to a perfectly
conducting plate $P_1$ in vacuum, and
a semi-infinite TI located at a
distance $a$. Here the plate and the
TI exert a force upon each
other. The Casimir energy in Eq. 
(\ref{CasEnergy}) in the limit $L
\rightarrow \infty$ takes the form 
$\mathcal{E}_{\theta} ^{L \rightarrow
\infty} = {\mathcal{E}}_a R ( \theta)$, 
with ${\mathcal{E}}_a= - \pi^{2} / 720 a ^{3}$, and the function
\begin{equation}
R (\theta) = \frac{120}{\pi ^{4}} \int _{0} ^{\infty} \xi ^{3} \frac{\tilde{\theta} ^{2}}{1 + \tilde{\theta} ^{2} e ^{- 2 \xi } \sinh ^{2} \xi} e ^{- 3 \xi } \sinh \xi d \xi ,
\end{equation}
is $a$-independent  and  bounded
by its  $\theta \rightarrow
\pm \infty$ limit, \textit{i.e.,}
\begin{equation}
R (\theta ) \leq \frac{120}{\pi ^{4}} \int _{0} ^{\infty} \xi ^{3} \frac{e ^{ - \xi }}{\sinh \xi}  d \xi = 1 .
\end{equation}
Thus, for this case,  the energy stored
in the electromagnetic field
is bounded
by the Casimir energy
between two parallel conducting plates at a distance $a$,
\textit{i.e.,} 
$\mathcal{E} _{\theta} ^{L \rightarrow
\infty} \leq {\mathcal{E}}_a$. Physically this
implies that in the  $\theta \rightarrow
\infty $ limit the surface of the TI
mimics a conducting plate, which is analogous to
Schwinger's prescription for describing a
conducting plate as the  $
\varepsilon \rightarrow \infty$ limit of 
material media \cite{CED}. 
{These  results, which stem from our
Eqs.~(\ref{theta-GF}) and (\ref{A(Z,Z)}), 
agree 
with those obtained in the 
%the scattering approach to compute the 
global 
energy approach which uses the reflection matrices 
containing the Fresnel
coefficients as in Ref.~\citenum{Grushin}, 
when 
the 
appropriate limits 
to describe an ideal conductor at $P_1$ and a purely topological surface at
 $\Sigma$ are taken 
into account. } 
Taking
the derivative with respect to $a$ we find
that the plate and the TI
exert a force (in units of $F_a = - \pi ^{2}
/ 240 a ^{4}$) of attraction upon each
other
given by $f _{\theta} = F ^{L \rightarrow
\infty} _{\theta} / F_a = R ( \theta )$. 
Numerical results for  $f_{\theta}$ for different values of $\theta$ are presented in Table \ref{table1}.

\begin{table}
\tbl{Normalized force $f_{\theta}=F ^{L \rightarrow
\infty} _{\theta} / F_a = R ( \theta )$ for different values of $\theta$.}
{\begin{tabular}{cccccc}
\toprule
$\theta$ & $\pm 7 \pi$ & $\pm 15 \pi$ & $\pm 23 \pi$ & $\pm 31 \pi$ & $\pm 39 \pi$
\\\colrule
$f _{\theta}$ & 0.0005 & 0.0025 & 0.0060 & 0.0109 & 0.0172 \\
\\\botrule
\end{tabular}}
\label{table1}
\end{table}
A general feature of our analysis is that the  TI
induces a $\theta$-dependence on the Casimir stress, which 
could be used to measure $\theta$. 
Since  the Casimir stress  has been measured for 
separation distances 
in the $0.5- 3.0\,\mu$m range \cite{Bressi}, these 
measurements  require
TIs of width lesser than $0.5 \mu$m and an increase of the  experimental precision of two to three orders of magnitude.
In  practice the ability to measure  $f_\theta$ 
depends on the value of the topological
MEP, which is
quantized as
$\theta = (2n + 1) \pi, \, n \in
\mathbb{Z}$.
The particular values $\theta = \pm 7 \pi
, \pm 15 \pi$ are appropriate for the TIs
such as Bi$_{1-x}$Se$_{x}$ 
{\cite{Zhou}}, where
we have  $f _{\pm 7 \pi}
\approx 0.0005$ and $f _{\pm 15 \pi}
\approx 0.0025$, which are not yet
feasible with the present experimental
precision. 
This
effect could also be explored in  TIs
described by a higher  coupling $\theta$, 
such as Cr$_{2}$O$_{3}$. 
However, this
material induces more
general magnetoelectric couplings   not
considered in 
our model \cite{Grushin}.

Although the reported $\theta$-effects of our Casimir systems cannot be  observed 
in the  laboratory yet, 
we have aimed
to establish the Green's function method as an alternative  theoretical framework
for dealing with the topological magnetoelectric effect of TIs
and also as yet another application of the GF method we
developed in Ref.~\citenum{AMU1,AMU2,AMU3}.

%%%%%%%%%%%%%%%%%%%%%%%%%%%%%%%

\section*{Acknowledgments}
LFU takes the opportunity to thank the organizers of the Julian Schwinger Centennial Conference and Workshop for a wonderful meeting honoring the great physicist and scholar. LFU also acknowledges support from the CONACyT (M\'exico) project No. 237503.  AM was supported by the CONACyT postdoctoral Grant No. 234774.

%to 12 February 2018. This event is jointly organized
%Non BiBTeX users can list down their references as:


\begin{thebibliography}{99}

\bibitem{MAXWELL} J. C. Maxwell, A Dynamical Theory of the Electromagnetic Field, \textit{Phil. Trans. R. Soc. London} \textbf{155}, 459 (1865).

\bibitem{SCHW1} J. S. Schwinger, On Quantum-Electrodynamics and the Magnetic Moment of the Electron, \textit{Phys. Rev.} \textbf{73}, 416 (1948).

\bibitem{SCHW2} J. S. Schwinger, Quantum Electrodynamics I. A covariant Formulation, \textit{Phys. Rev.} \textbf{74}, 1439 (1948).

\bibitem{SCHW3} J. S. Schwinger, Quantum Electrodynamics II. Vacuum Polarization and Self Energy, \textit{Phys. Rev.} \textbf{75}, 651 (1949).

\bibitem{PONT} C. Nash and S. Sen, \textit{Topology and Geometry for
Physicists} (London: Academic Press Inc, 1983.)

\bibitem{ANOM} K. Fujikawa and H. Suzuki, \textit{Path Integrals and Quantum
Anomalies} (Oxford: Clarendon Press, 2004).

\bibitem{SCP} M. Dine, TASI Lectures on the Strong CP Problem arXiv:0011376
[hep-ph],2000.

\bibitem{TFT} D. Birmingham, M. Blau, M. Radowski and G. Thompson,
Topological field theory \textit{Phys. Rep.} \textbf{209}, 129 (1991).

\bibitem{AXIONS} M. Kuster, G. Raffelt and B. Beltr\'{a}n (eds), \textit{%
Axions: Theory, Cosmology, and Experimental Searches} (\textit{Lecture Notes
in Physics} vol 741) (Berlin: Springer-Verlag, 2008).

\bibitem{TI} X. L. Qi , T. L. Hughes and S. C. Zhang, Topological field
theory of time-reversal invariant insulators, \textit{Phys. Rev.} B \textbf{78%
}, 195424 (2008).

\bibitem{LODELL} T. H. O'Dell, \textit{The Electrodynamics of Magneto-Electric Media} (North-Holland, Amsterdam, 1970);  L. D. Landau, E. M. Lifshitz and L. P. Pitaevskii, 
\textit{Electrodynamics of Continuous Media} (\textit{Course of Theoretical
Physics} vol 8) (Oxford: Pergamon Press, 1984).

\bibitem{MetaMat} E. Plum, J. Zhou, J. Dong, V. A. Fedotov, T. Koschny, C.
M. Soukoulis and N. I. Zheludev, Metamaterial with negative index due to
chirality, \textit{Phys. Rev.} B \textbf{79}, 035407 (2009).


\bibitem{WS} M. M. Vazifeh and M. Franz, Electromagnetic Response of Weyl
Semimetals, \textit{Phys. Rev. Lett.} \textbf{111}, 027201 (2013).

\bibitem{TI1} X. L. Qi, \textit{Field-Theory Foundations of Topological Insulators}, in \textit{Topological Insulators (Contemporary Concepts of Condensed Matter Science)}, eds.M. Franz and L. Molenkamp  , Vol. 6 (Amsterdam: Elsevier, 2013).

\bibitem{WS1} N. P. Armitage, E. J. Mele and A. Vishwanath, Weyl and Dirac
semimetals in three-dimensional solids, \textit{Rev. Mod. Phys.} \textbf{90},
015001 (2018).

\bibitem{DZYA} I. E. Dzyaloshinskii, On the Magneto-Electrical Effect in Antiferromagnets, JETP \textbf{37}, 881 (1959).

\bibitem{ASTROV} D. N. Astrov, The Magneto-Electrical Effect in Antiferromagnets, JETP \textbf{38}, 984 (1960).

\bibitem{FIEBIG} M. Fiebig, Revival of the magnetoelectric effect, J. Phys. D: Appl. Phys. \textbf{38}, R123 (2005).

\bibitem{METI} V. Diziom, A. Shuvaev, A. Pimentov et al., Observation of the universal magnetoelectric effect in a 3D topological insulator, \textit{Nature Communications} \textbf{8}, 15297 (2017).


\bibitem{science} X. L. Qi, R. Li, J. Zang and S. C. Zhang, Inducing a
magnetic monopole with topological surface States, \textit{Science} \textbf{%
323}, 1184 (2009).


\bibitem{Kim1} C. Kim, E. Koh, and K. Lee, Janus and Multifaced Supersymmetric Theories, \textit{Journal of High Energy Physics}, 
\textbf{0806}, 040 (2008).

\bibitem{Kim2} C. Kim, E. Koh, and K. Lee, Janus and multifaced supersymmetric theories. II, \textit{Phys. Rev. D} \textbf{79}, 126013
(2009).

\bibitem{Wilczek} F. Wilczek, Two applications of axion electrodynamics, 
\textit{Phys. Rev. Lett.} \textbf{58}, 1799 (1987).


\bibitem{ZH}  L. Huerta and J. Zanelli, Optical properties of a $\theta$ vacuum, \textit{Phys. Rev. D} \textbf{85}, 085024
(2012).



\bibitem{Hehl} Y. N. Obukhov and F. W. Hehl, Measuring a piecewise constant axion field in classical electrodynamics, \textit{Phys. Lett. A} \textbf{341}, 357
(2005).

\bibitem{PRA} A. Mart\'\i n-Ruiz and L. F. Urrutia, Interaction of a
hydrogenlike ion with a planar topological insulator, \textit{Phys. Rev.} A 
\textbf{97}, 022502 (2018).

\bibitem{EPL} A. Mart\'{\i}n-Ruiz and E. Chan-L\'{o}pez, Dynamics of a
Rydberg hydrogen atom near a topologically insulating surface, \textit{Eur.
Phys. Lett.} \textbf{119}, 53001 (2017)

\bibitem{Jackson} J. D. Jackson, \textit{Classical Electrodynamics} (Hoboken
NJ: John Wiley \& Sons, 1999).



\bibitem{CED} J. Schwinger, L. DeRaad, K. Milton and W. Tsai, \textit{%
Classical Electrodynamics}, Advanced Book Program, (Perseus Books 1998).


\bibitem{EPL2} A. Mart\'{\i}n-Ruiz, M. Cambiaso and L. F. Urrutia, A Green's
function approach to the Casimir effect on topological insulators with
planar symmetry, \textit{Eur. Phys. Lett.} \textbf{113}, 60005 (2016).



\bibitem{AMU1} A. Mart\'\i n-Ruiz, M. Cambiaso and L. F. Urrutia,
A Green's function approach to Chern-Simons extended electrodynamics: An
effective theory describing topological insulators, \textit{Phys. Rev. D} \textbf{92}, 125015 (2015).


\bibitem{AMU2} A. Mart\'{\i}n-Ruiz, M. Cambiaso and L. F. Urrutia,
Electro- and magnetostatics of topological insulators as modeled by planar,
spherical, and cylindrical $\theta$ boundaries: Green's function approach 
\textit{Phys. Rev.} D \textbf{93}, 045022 (2016).

\bibitem{AMU3} A. Mart\'\i n-Ruiz, M. Cambiaso and L. F. Urrutia,
Electromagnetic description of three-dimensional time-reversal invariant
ponderable topological insulators, \textit{Phys. Rev.} D \textbf{94}, 085019
(2016).

%%%%%CASIMIR EFFECT

\bibitem{Casimir} H. B. G. Casimir, On the attraction between two perfectly conducting plates, \textit{Proc. K. Ned. Akad. Wet.} \textbf{51}, 793 (1948).

\bibitem{Bressi} G. Bressi \textit{et al}, Measurement of the Casimir Force between Parallel Metallic Surfaces, \textit{Phys. Rev. Lett.} \textbf{88}, 041804 (2002).

\bibitem{Milton} K. A. Milton, \textit{The Casimir effect: Physical Manifestation of Zero-Point Energy} (World Scientific, Singapore, 2001).

\bibitem{Bordag} M. Bordag, G. L. Klimchitskaya, U. Mohideen and V. M. Mostepanenko, \textit{Advances in Casimir effect} (Oxford University Press, Great Britain, 2009).


\bibitem{Liang} L. Fu, C. L. Kane, and E. J. Mele, Topological Insulators in Three Dimensions, \textit{Phys. Rev. Lett.} \textbf{98}, 106803 (2007); D. Hsieh, D. Qian, L. Wray, Y. Xia, Y. S. Hor, R. J. Cava
and M. Z. Hasan, A topological Dirac insulator in a quantum spin Hall phase, \textit{Nature} \textbf{452}, 970 (2008).

\bibitem{Grushin} A. G. Grushin and A. Cortijo, Tunable Casimir Repulsion with Three-Dimensional Topological Insulators, \textit{Phys. Rev. Lett.} \textbf{106}, 020403 (2011); {A. G. Grushin, P. Rodriguez-Lopez and A. Cortijo, Effect of finite temperature and uniaxial anisotropy on the Casimir effect with three-dimensional topological insulators, \textit{Phys. Rev. B} \textbf{84}, 045119 (2011).}


\bibitem{Brown} L. S. Brown and G. J. Maclay, Vacuum Stress between Conducting Plates: An Image Solution, \textit{Phys. Rev.} \textbf{184}, 1272 (1969).



\bibitem{Schwinger1} J. Schwinger, L. DeRaad and K. Milton, Casimir effect in dielectrics, \textit{Ann. Phys. (N.Y.)} \textbf{115}, 1 (1978).



\bibitem{Candelas} D.~Deutsch and P.~Candelas, Boundary effects in quantum field theory, \textit{Phys. Rev. D} \textbf{20}, 3063 (1979).

\bibitem{Cavalcanti} R. M. Cavalcanti, Casimir force on a piston, \textit{Phys. Rev. D} \textbf{69}, 065015 (2004).


 
\bibitem{Zhou} X.~Zhou, \textit{et al}, Photonic spin Hall effect in topological insulators, \textit{Phys. Rev. A} \textbf{88}, 053840 (2013).
%  doi:10.1103/PhysRevA.88.053840
  

\end{thebibliography}
\end{document}